\documentclass[iop,numberedappendix]{emulateapj}
\usepackage{epsfig}
\usepackage{amsmath}
\usepackage{enumerate}
\usepackage{natbib}
\usepackage{hyperref}

\usepackage{ulem}
\usepackage[squaren, Gray, cdot]{SIunits}

\usepackage{color}

\shorttitle{Non-thermal X-rays from shocks in novae}
\shortauthors{Vurm \& Metzger}

\def\me{m_{\rm e}}
\def\mprot{m_{\rm p}}

\def\nrad{n_{\rm rad}}

\def\Qe{Q_{\rm e}}
\def\Qp{Q_{\rm p}}

\def\Qinj{Q_{\rm inj}}

\def\epsB{\varepsilon_B}
\def\epsnth{\varepsilon_{\rm nth}}

\def\d{\delta}

\def\pth{p_{\rm th}}

\def\ath{\alpha_{\rm th}}

\def\pnth{p_{\rm nth}}
\def\unth{u_{\rm nth}}
\def\pb{p_B}
\def\ub{u_B}
\def\hb{h_B}
\def\ab{\alpha_B}

\def\pprot{p_{\rm p}}
\def\uprot{u_{\rm p}}

\def\vw{v_{\rm w}}
\def\vsh{v_{\rm sh}}
\def\vds{v_{\rm ds}}
\def\nds{n_{\rm ds}}
\def\texp{t_{\rm exp}}

\def\tcl{t_{\rm line}}
\def\tcbr{t_{\rm br}}
\def\tcIC{t_{\rm IC}}
\def\tcCoul{t_{\rm Coul}}

\def\tcpp{t_{\rm pp}}

\def\dg{\dot{\gamma}}
\def\dgbr{\dot{\gamma}_{\rm br}}
\def\dgC{\dot{\gamma}_{\rm Coul}}
\def\dgIC{\dot{\gamma}_{\rm IC}}
\def\dgsyn{\dot{\gamma}_{\rm syn}}

\def\chisyn{\chi_B}
\def\uB{u_B}

\def\sigmaT{\sigma_{\rm T}}
\def\spp{\sigma_{\rm pp}}
\def\afs{\alpha_{\rm fs}}

\def\tauT{\tau_{\rm T}}
\def\lrad{l_{\rm rad}}

\def\urad{u_{\rm opt}}
\def\Lrad{L_{\rm opt}}
\def\Lg{L_{\gamma}}
\def\Lsh{L_{\rm shock}}

\def\epm{e_{\pm}}

\def\nue{\nu_{\rm e}}
\def\numu{\nu_{\mu}}
\def\anue{\overline{\nu}_{\rm e}}
\def\anumu{\overline{\nu}_{\mu}}

\def\me{m_{\rm e}}

\def\jbrep{j_{\rm br, ep}}
\def\jIC{j_{\rm IC}}
\def\xopt{\overline{x}_{\rm opt}}
\def\Eopt{\overline{E}_{\rm opt}}

\def\gmin{\gamma_{\rm min}}
\def\gmax{\tilde{\gamma}}

\def\epsp{\varepsilon_{\rm p}}
\def\Np{N_{\rm p}}
\def\Ep{E_{\rm p}}
\def\gp{\gamma_{\rm p}}
\def\bp{\beta_{\rm p}}
\def\qp{q_{\rm p}}
\def\gpmax{\gamma_{\rm p, max}}

\def\Mdot{\dot{M}}

\def\ba{\begin{align}}
\def\ea{\end{align}}

\newbox\grsign \setbox\grsign=\hbox{$>$} \newdimen\grdimen \grdimen=\ht\grsign
\newbox\simlessbox \newbox\simgreatbox \newbox\simpropbox
\setbox\simgreatbox=\hbox{\raise.5ex\hbox{$>$}\llap
     {\lower.5ex\hbox{$\sim$}}}\ht1=\grdimen\dp1=0pt
\setbox\simlessbox=\hbox{\raise.5ex\hbox{$<$}\llap
     {\lower.5ex\hbox{$\sim$}}}\ht2=\grdimen\dp2=0pt
\setbox\simpropbox=\hbox{\raise.5ex\hbox{$\propto$}\llap
     {\lower.5ex\hbox{$\sim$}}}\ht2=\grdimen\dp2=0pt

\begin{document}

\title{High-energy emission from non-relativistic radiative shocks: application to gamma-ray novae} 

\author{
Indrek Vurm\altaffilmark{1,2}, Brian D.~Metzger\altaffilmark{1}
}
\affil{$^1$Physics Department and Columbia Astrophysics Laboratory, Columbia University, 538 West 120th Street, New York, NY 10027, USA; indrek.vurm@gmail.com \\
	$^2$Tartu Observatory, T$\tilde{o}$ravere, Tartumaa EE-61602, Estonia\\
}

\label{firstpage}
\begin{abstract}

Multiwavelength radiation from relativistic particles accelerated at shocks in novae and other astrophysical sources
carries a wealth of information about the outflow properties and the microphysical processes at work near the shocks.
The observation of GeV gamma-rays from novae by {\it Fermi}/LAT demonstrates that the shocks in these systems
can accelerate particles to energies of at least $\sim 10$~GeV.  The low-energy extension of the same
non-thermal particle distribution inevitably gives rise to emission extending into the X-ray band.
Above $\gtrsim 10$~keV this radiation can escape the system without significant absorption/attenuation,
and can potentially be detected by {\it NuSTAR}.  We present theoretical models for
hard X-ray and gamma-ray emission from radiative shocks in both leptonic and hadronic scenarios,
accounting for the rapid evolution of the downstream properties due to the fast cooling of thermal plasma.
Due to strong Coulomb cooling of
the mildly relativistic electrons nominally 
responsible for producing hard X-ray emission, only a fraction of $10^{-4} - 10^{-3}$ of the gamma-ray luminosity
is radiated in the {\it NuSTAR} band; nevertheless, this emission could be detectable simultaneous with
the LAT emission in bright gamma-ray novae with a $\sim 50$~ks exposure.
The spectral slope in hard X-rays is
$\alpha \approx 0$ for typical nova parameters,
thus serving as a testable prediction of the model.  Our work demonstrates how combined hard X-ray and gamma-ray observations
can be used to constrain properties of the nova outflow (velocity, density and mass outflow rate) and particle acceleration at the shock.
A very low X-ray to gamma-ray luminosity ratio ($L_{\rm X}/L_{\gamma} \lesssim 5 \times 10^{-4}$)
would disfavor leptonic models for the gamma-ray emission.  Our model can also be applied
to other astrophysical environments with radiative shocks, including Type IIn supernovae and colliding winds in massive star binaries.

\end{abstract}



\section{Introduction}

There are multiple lines of evidence that shocks play an important role in shaping the
multiwavelength emission from nova outflows.
Among these are the peculiar early-time radio light curves exhibiting significantly higher brightness temperatures
than can be explained by passively cooling expanding outflows (\citealt{Krauss+11}; \citealt{Metzger+14}; \citealt{Weston+15}; \citealt{Yang+15}, \citealt{Vlasov+16}).
Among the most 
compelling evidence to date is the unexpected detection of GeV gamma-rays from three
classical novae by {\it Fermi}/LAT in 2012-2013:
V959 Monocerotis, V1324 Scorpii and V339 Delphini (\citealt{Ackermann+14});
since then, at least 2 more novae, V1369 Cen and V5668 Sgr, have been detected with high significance (\citealt{Cheung+16}).  The comparatively high GeV luminosities of novae suggest that
shocks dissipate
a sizable fraction of the total energy budget, and could in some cases even dominate the optical output \citep{Metzger+15}.

Despite clear evidence for shocks, their origin and location within the nova outflow remains uncertain.
\citet{Chomiuk+14} suggests a connection between the shocks and the bipolar geometry of the nova outflow (e.g.~\citealt{Shore13}, \citealt{Ribeiro+13}),
which in turn could be shaped by the gravity of the orbiting binary companion (e.g.~\citealt{Livio+90}).
Unfortunately, this is challenging to confirm because the shock emission cannot be resolved when the gamma-rays are detected near the peak of the nova outburst.

Compared to the vast range of different astrophysical sources where shocks play a significant role,
the shocks in nova outflows probe a fairly unique regime in parameter space,
characterized by
relatively
high densities and low velocities.
In this regime the shocks giving rise to detectable gamma-rays are likely to be radiatively efficient,
i.e. all of the dissipated energy is radiated as multiwavelength emission.

Several works have been dedicated to calculating the shock emission in different frequency bands,
ranging from radio to TeV gamma-rays (\citealt{OBrien+94}, \citealt{Nelson+12}, \citealt{Martin&Dubus13}, \citealt{Metzger+14}, \citealt{Metzger+15}, \citealt{Metzger+16}, \citealt{Vlasov+16}).
In this work we focus on the hard X-ray band ($\gtrsim$ 10 keV) accessible to e.g.~{\it NuSTAR},
and explore in particular the connection between the radiation properties in the X-ray and gamma-ray bands.
Our work is motivated by recent {\it NuSTAR} observations of V339 Del and V5668 Sgr,
which place rather stringent upper-limits on the hard X-ray emission simultaneous with
the LAT detections (Mukai et al., in prep); within the framework of our model,
these and future observations can be used to place constraints on the location and the electron/ion acceleration efficiencies of the shocks in novae.  

Although our analysis is focused primarily on shocks in novae, our results are also applicable to other astrophysical sources of non-relativistic radiative shocks.  These include, for instance, the dense colliding stellar winds of massive binary stars (e.g., \citealt{DeBecker07}) and Type IIn supernovae, in which the supernova ejecta collides with a dense external shell of gas surrounding the progenitor star (e.g.,~\citealt{Chevalier&Fransson94}, \citealt{Smith+07}).

\subsection{Non-relativistic shocks in dense media}

Shocks taking place in e.g. gamma-ray novae ($\vsh \approx 10^8$~cm~s$^{-1}$) and Type IIn supernovae ($\vsh \approx 10^9$~cm~s$^{-1}$)
heat the bulk of the gas to X-ray temperatures.
In sufficiently dense media, the cooling time of the shocked gas is short compared to the dynamical time of the system.  This results in strong compression of the gas in the shock downstream as it cools (e.g., \citealt{Drake05}).

Strong observational evidence also exists that non-relativistic shocks can accelerate particles (either electrons and/or protons/ions) to ultrarelativistic energies,
which emit broadband non-thermal radiation from radio to gamma-ray frequencies.
However, in contrast to relativistic shocks, in non-relativistic radiative shocks
the cooling time of the relativistic particles can exceed the cooling/compression time of the thermal gas behind the shock.
This has two effects on the non-thermal particles and their radiation:
(1) the rising density in the downstream alters the relative importance of different radiative processes,
most importantly relativistic bremsstrahlung and inverse Compton (IC) emission, 
and (2) rapid adiabatic compression supplies additional energy to the non-thermal particles.
Thermal cooling thus affects both the radiative efficiency and spectral shape of the non-thermal emission.

\subsection{Radiative processes in leptonic and hadronic scenarios}

The main goal of this paper is to establish a theoretical framework which enables one to combine non-thermal X-ray and gamma-ray data into a diagnostic tool for the shock environment and the properties of non-thermal particle acceleration.  Regardless of whether leptonic or hadronic processes are responsible for the gamma-ray emission, the same radiative processes inevitably also gives rise to X-ray radiation.  The relative luminosity in the X-ray and gamma-ray bands, $L_X/L_{\gamma}$ is most sensitive to the ratio of matter to radiation energy density,
which in turn depends on the density of the shocked gas and the location of the shock within the nova outflow.
The ratio $L_X/L_{\gamma}$ is also sensitive to whether hadrons or leptons dominate the accelerated non-thermal particle populations and on the injected particle spectra.

In the leptonic scenario,
gamma-ray emission is the result of direct electron acceleration;
the dominant non-thermal radiative processes are relativistic bremsstrahlung and IC scattering.
The observed gamma-ray luminosities require the injected energy spectrum to be almost logarithmically flat, i.e. $q=2$ (where $dN_{\rm inj}/d\gamma \propto \gamma^{-q}$ and $\gamma$ is the particle Lorentz factor),
to avoid an energy crisis (\citealt{Metzger+15}).
This is consistent with the approximately flat $\nu F_{\nu}$ spectra observed in gamma-ray novae (\citealt{Ackermann+14}).
If $q=2$, the bremsstrahlung and IC spectra are similarly flat in the gamma-ray range,
thus limiting the diagnostic value of the gamma-ray spectrum alone in determining the density and radiation compactness of the shock.  Non-thermal hard X-ray emission provides an independent diagnostic, which is comparatively more luminous at low densities and high (optical) luminosities, for which IC cooling dominates the non-thermal emission.

The high GeV luminosities of LAT-detected novae require shocks to occur in relatively dense environments.
Non-thermal X-ray emission in such cases results from a combination of bremsstrahlung and IC emission,
modified by Coulomb (and possibly synchrotron) losses.  Indeed, we will show that electrons with
Lorentz factors $\gamma \lesssim 10^3$ lose most of their energy via Coulomb collisions with the thermal population,
which significanly (though not completely) suppresses their radiative output below the LAT band.
As a result, the gamma-ray spectrum breaks to a steeper slope at energies below a few hundred MeV;
a naive extrapolation of the LAT spectrum to the {\it NuSTAR} band would therefore grossly overestimate the X-ray flux, by as much as three orders of magnitude.

In the hadronic scenario, when the energy in accelerated protons dominates over electrons,
the gamma-rays are mainly generated by the production and decay of neutral pions ($\pi_{0}$), created by proton-proton/ion collisions.
These collisions also produce charged pions ($\pi_{\pm}$) that ultimately decay into relativistic electron-positron pairs carrying energy comparable to that in gamma-rays from $\pi_0$ decay.
As in the leptonic model, the created pairs radiate both X-rays and gamma-rays via bremsstrahlung and IC emission, which dominates the emission between $\sim 10$~keV and $\sim 100$~MeV.
The main difference from the leptonic scenario is the paucity of injected pairs with energies well below the pion rest mass $\sim 100$~MeV,
which would otherwise make a significant contribution to the X-ray flux (despite Coulomb losses).
Overall,
the additional $\pi_0$ gamma rays,
coupled with fewer X-ray emitting leptons, result in systematically lower ratio of X-ray to gamma-ray flux in the hadronic scenario.

This paper is organized as follows.
In Section \ref{sec:rad_mech} we give an overview of the radiative processes relevant
in nova shocks.
The evolution of the heated plasma in the downstream of radiative shocks
is discussed in Section \ref{sec:gasdyn}.
In Section \ref{sec:nthevo} we describe
the evolution of non-thermal particle distributions
as they radiate and cool in the compressing downstream flow.
A theoretical overview of the X- and gamma-ray emission from the cooling layer is given in 
Section \ref{sec:emission}.
The numerical results and 
the constraints on parameter space from simultaneous hard X-ray and gamma-ray observations
are presented in Sections \ref{sec:Xgratio} and \ref{sec:Mdotn}.
Our results are discussed and conclusions summarized in Section \ref{sec:concl}.

\section{Radiative mechanisms and post-shock cooling}

\label{sec:rad_mech}

\subsection{Thermal processes}

Consider the plasma downstream of a non-relativistic shock.
The bulk of the shock energy
is transferred to thermal plasma, which provides the pressure support in the immediate downstream.
At shock velocities $\vsh \sim 10^8$~cm~s$^{-1}$
the post-shock temperature corresponds to soft X-rays ($T \approx 1.7\times 10^7 v_{{\rm sh},8}^2$,
where $v_{\rm sh,8} \equiv v_{\rm sh}/10^{8}$~cm~s$^{-1}$).
The thermal plasma cools via free-free emission and line cooling.
The latter dominates at
$T\lesssim 10^8$~K;
the shock is thus radiative if $\tcl < \texp=R/v_{\rm sh}$, where
\begin{align}
\tcl &\approx \frac{3k T}{8\mu n \Lambda_{\rm line}} \approx 3.1\times 10^3 \, T_7^{1.7} n_{9}^{-1} \, \mbox{s} \nonumber \\
&\approx 7.8\times 10^3 \, v_{{\rm sh},8}^{3.4} \, n_{9}^{-1} \, \mbox{s},
\label{eq:cool:l}
\end{align}
$\mu = 0.76$ is the mean molecular weight appropriate for nova composition (\citealt{Schwarz+07}, \citealt{Vlasov+16}), $n = 10^{9}n_9$ cm$^{-3}$ is the upstream density,
and we have approximated $\Lambda_{\rm line} \approx 2.2\times 10^{-22} \, (T/10^7)^{-0.7}$~erg~cm$^3$~s$^{-1}$
(\citealt{Schure+09}; see also \citealt{Vlasov+16}).  On a week timescale relevant for gamma-ray emission, the shock is likely to be radiative if $\vsh \lesssim 2\times 10^8$~cm~s$^{-1}$ unless the density is very low, which however would result in a gamma-ray luminosity too low to be detected by {\it Fermi} (\citealt{Metzger+15}).

If the shock is indeed radiative, cooling of the thermal plasma leads to strong compression in the downstream in order to maintain the required pressure.
This compression is halted only once either non-thermal or magnetic pressure becomes dominant over thermal pressure,
or once the gas cools to temperatures $\lesssim 10^4$~K below which line cooling becomes less efficient due to recombination of the gas.

The luminous thermal $\sim$ keV X-rays from the gamma-ray emitting shocks are not directly observable, as they are absorbed by bound-free processes in the material ahead of the shock.\footnote{
Some novae show hard $\gtrsim$ keV thermal X-ray emission of luminosity $L_X \sim 10^{32}-10^{35}$ erg s$^{-1}$
within days to weeks of the ouburst (e.g. V5589 Sgr; \citealt{Weston+15b}), consistent with being powered
by adiabatic (non-radiative) shocks (\citealt{Mukai&Ishida01}; \citealt{Osborne15}).
However, the kinetic power of these `fast' X-ray producing shocks are generally too low to explain the luminous LAT GeV emission,
suggesting that they originate from a different location within the ejecta (e.g.~\citealt{Vlasov+16}). }
Instead, this energy is reprocessed to lower frequencies and released as optical/UV radiation (\citealt{Metzger+14}).
In contrast, non-thermal X-rays with higher energy $\gtrsim 10$~keV are not significantly attenuated by bound-free absorption,
and only interact with the ejecta via Compton (Thomson) scattering.  They could therefore potentially be detected at early times simultaneously with the gamma-ray emission.

\subsection{Non-thermal processes: leptonic scenario}

\label{sec:nth:lep}

In addition to thermal heating, a portion of the shock energy is used to accelerate
a fraction of the electrons and/or baryons into a non-thermal distribution.
In the leptonic scenario, the relativistic electrons cool via
IC emission on (primarily) optical/UV photons, 
relativistic bremsstrahlung emission, Coulomb collisions with thermal electrons,
and synchrotron emission if the downstream is appreciably magnetized.
The interplay between these processes determines both the dominant radiative mechanism at hard X- and gamma-ray frequencies,
as well as the partitioning of the non-thermal luminosity between different bands.

The key parameter that determines the dominant cooling mechanism of relativistic electrons
is the ratio of soft (optical) radiation energy density (that determines the IC cooling rate)
to matter density (determines bremsstrahlung and Coulomb losses)\footnote{Equivalently,
$\chi$ can be defined as the ratio of radiation compactness $\lrad = \sigmaT \urad R/(\me c^2)$
to the Thomson opacity $\tauT = \sigmaT n R$, $\chi = \lrad/\tauT$.},
\begin{align}
\chi \equiv \frac{\urad}{m_e c^2 n}.
\label{eq:chi}
\end{align}
For typical parameters in gamma-ray novae one obtains
\begin{align}
\chi =
3.2\times 10^{-5} \, \frac{L_{{\rm opt}, 38}}{n_9 \, R^2_{14}},
\label{eq:chichar}
\end{align}
where $\Lrad = 10^{38}L_{\rm opt,38}$ erg s$^{-1}$ is the optical luminosity,  $R = 10^{14}R_{14}$ cm is the shock radius.
We have used $\urad = \Lrad/(4\pi c R^2)$, i.e.~neglecting the $(1+\tauT)$ correction  under the assumption that
the Thomson optical depth $\tauT \simeq n R \sigma_T \approx 0.06 n_9 R_{14}$ of the shocks is $\lesssim 1$.  

If the bulk of the optical luminosity is generated by reprocessed emission from the shock itself, then
\begin{align}
\Lrad \approx \Lsh &= \frac{9\pi}{8} m_p n \vsh^3 R^2 f_{\Omega} 	\nonumber \\
&= 5.9\times 10^{37} \, R_{14}^2 \, n_{9} \, v_{{\rm sh}, 8}^3  f_{\Omega} \,\, \mbox{erg s}^{-1}
\label{eq:Lsh}
\end{align}
where $f_{\Omega}$ is the fraction of the total solid angle subtended by the shock.  Again expressing $\urad$, one obtains from Equation (\ref{eq:chi})
\begin{align}
\chi = 1.9 \times 10^{-5} \, v_{{\rm sh},8}^3.
\label{eq:chi_min}
\end{align}
This represents the minimal value of $\chi$ that can be attained at a given shock speed.

Consider separately the cooling rates by different processes.  Free-free emission from electrons of energy $\gamma \gtrsim $~a~few
receives comparable contributions from electron-electron and electron-proton bremsstrahlung.
For analytical estimates we employ the approximate expression (accurate within $<15$~\% for $\gamma=10-10^4$)
\begin{align}
\dgbr
\approx \frac{5}{6} \, c\sigmaT\afs \nds \gamma^{1.2}  \sum_i \frac{X_i Z_i(1+Z_i)}{A_i}
\approx \frac{5}{3} \, c\sigmaT\afs \nds \gamma^{1.2},
\label{eq:gdotbr}
\end{align}
where $\nds \approx 4n=\rho/m_p$ is the {\it downstream} density of the shock, $\alpha_{\rm fs} \simeq 1/137$ is the fine structure constant,
and the sum is taken over the atomic species of mass fraction $X_i$, charge $Z_i$ and atomic weight $A_i$.
We use a more accurate expression valid in both relativistic and non-relativistic regimes given by \citet{Haug04} in our numerical calculations.

The IC cooling rate in the Thomson regime is
\begin{align}
\dgIC
= \frac{4\sigmaT \urad (\gamma\beta)^2}{3m_e c},
\label{eq:gdotIC}
\end{align}
where
$\beta=(1 - 1/\gamma^2)^{1/2}$.
Its ratio to the bremsstrahlung cooling rate is (in the $\gamma \gg 1$ limit)
\begin{align}
\frac{\dgIC}{\dgbr} =
\left(\frac{\gamma}{\gamma_{\star\star}}\right)^{0.8},
\end{align}
where
\begin{align}
  \gamma_{\star\star} &= \left( \frac{5\afs}{\chi} \right)^{1.25}	\nonumber \\
  &=
  \left\{\begin{array}{ll}
	6.6\times 10^3 \,\left( \frac{\displaystyle n_9 \, R_{14}^2}{\displaystyle L_{{\rm opt}, 38}} \right)^{1.25},	&	L_{\rm opt}>\Lsh \vspace{1mm}	\\
	1.3\times 10^4 \,\, v_{{\rm sh}, 8}^{-3.75},									&	L_{\rm opt} \approx \Lsh.
	\end{array}
  \right.
\label{eq:gss}
\end{align}
Here, the first case assumes that the optical luminosity is external to, and greater than, that generated at the shock.
In the second case the optical luminosity is given by Equation (\ref{eq:Lsh}).

The Coulomb cooling rate is
\begin{align}
\dgC = \frac{3}{2} \, \ln{\Lambda} \,\frac{c\sigmaT \nds}{\beta} \sum_i \frac{X_i Z_i}{A_i} \approx
\frac{3}{2} \, \ln{\Lambda} \,\frac{c\sigmaT \nds}{\beta}.
\label{eq:gdotC}
\end{align}
The ratio of bremsstrahlung and Coulomb cooling is independent of the shock parameters,
\begin{align}
\frac{\dgbr}{\dgC}
\approx \left(\frac{\gamma}{\gamma_{\star}}\right)^{1.2},
\end{align}
where
\begin{align}
\gamma_{\star} = \left(\frac{\ln{\Lambda}}{\afs}\right)^{0.83} \approx 900 \,\, \left(\frac{\ln{\Lambda}}{25}\right)^{0.83}
\label{eq:gstar}
\end{align}
denotes the electron energy below which
bremsstrahlung emission is affected by Coulomb losses.

Depending on the downstream magnetization, synchrotron radiation can also be significant.
Though synchrotron emission is unlikely to be detectable at early times when gamma rays are observed,
due to free-free absorption (\citealt{Metzger+14}),
it manifests indirectly by attenuating the power emitted in the LAT band.
In complete analogy with IC cooling, the ratio of synchrotron to bremsstrahlung cooling rates is
\begin{align}
\frac{\dgsyn}{\dgbr} =
\left(\frac{\gamma}{\gamma_{\dagger}}\right)^{0.8},
\end{align}
where
\begin{align}
  \gamma_{\dagger} &= \left( \frac{5\afs}{\chisyn} \right)^{1.25} = 1.8\times 10^5 \, \varepsilon_{B, -4}^{-1.25} \, v_{{\rm sh}, 8}^{-2.5},	
\label{eq:gdag}
\end{align}
and
\begin{align}
\chisyn = \frac{\uB}{\me c^2 n} = 2.3\times 10^{-6} \, \varepsilon_{B, -4} \, v_{{\rm sh}, 8}^2. 
\end{align}
Here $\epsB$ parametrizes the post-shock magnetic energy density in terms of the total energy density as $\uB=(9/8)\mprot\epsB n \vsh^2$. 

The above expressions for $\gamma_{\star\star}$ and $\gamma_{\dagger}$ are calculated for conditions immediately after the shock.
However, when the plasma compresses further downstream, the bremsstrahlung and Coulomb cooling rates are enchanced proportionally to $n$,
while the IC rate is unaffected by compression.  As a result, the Lorentz factor above which IC dominates over free-free cooling
increases as
$\gamma_{\star\star} \propto \chi^{-1.25} \propto n^{1.25}$.  Synchrotron losses are also enhanced by compression,
to a greater extent than bremsstrahlung: $\dgsyn/\dgbr \propto n^{\ab-1}$, where
$\uB \propto n^{\ab}$ and the adiabatic index $\ab = 4/3 - 2$ depending on the magnetic field configuration.

\begin{figure*} 
  \begin{center}
  \begin{tabular}{cc}
  \includegraphics[width=0.45\textwidth]{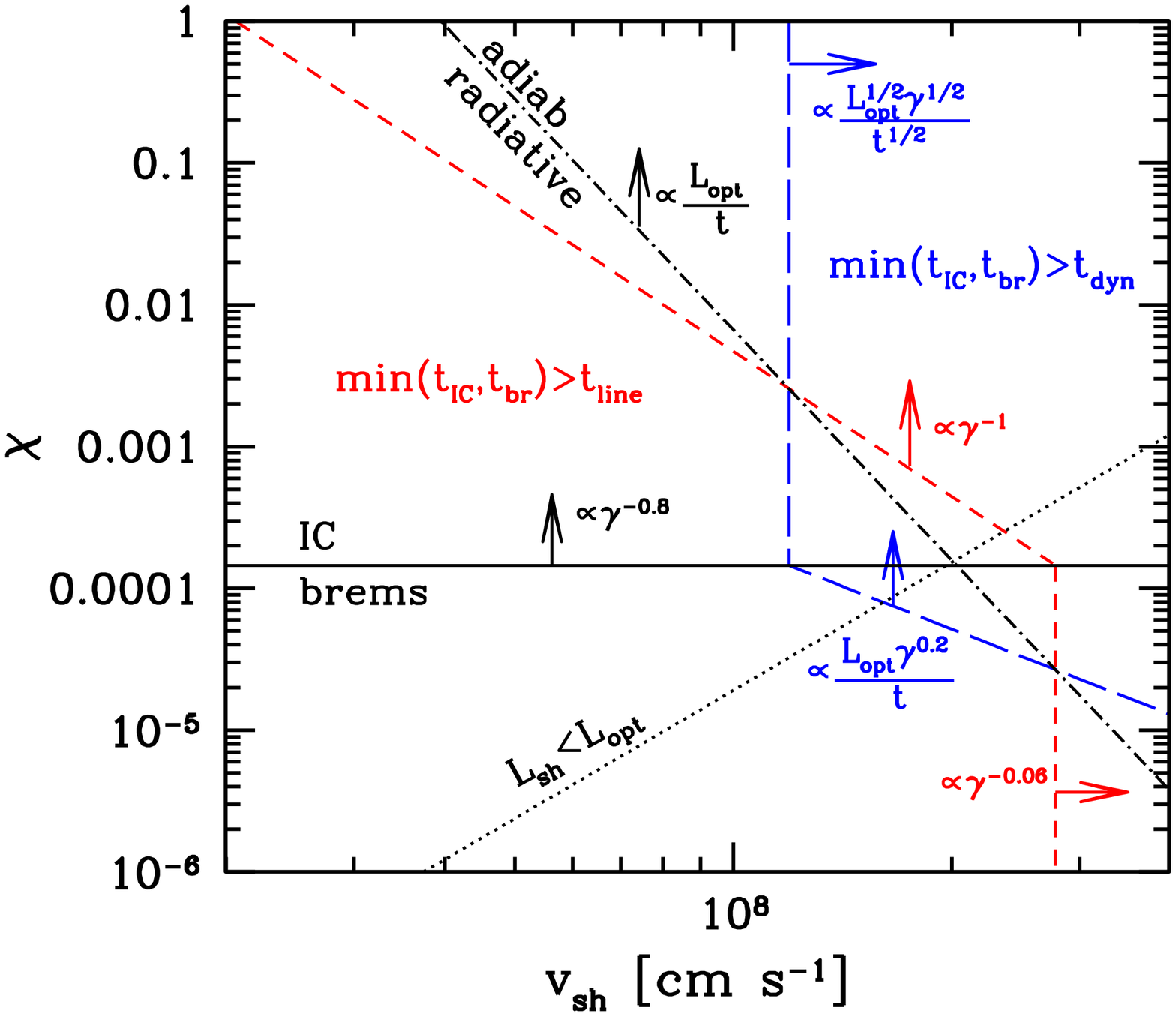}	&
  \includegraphics[width=0.45\textwidth]{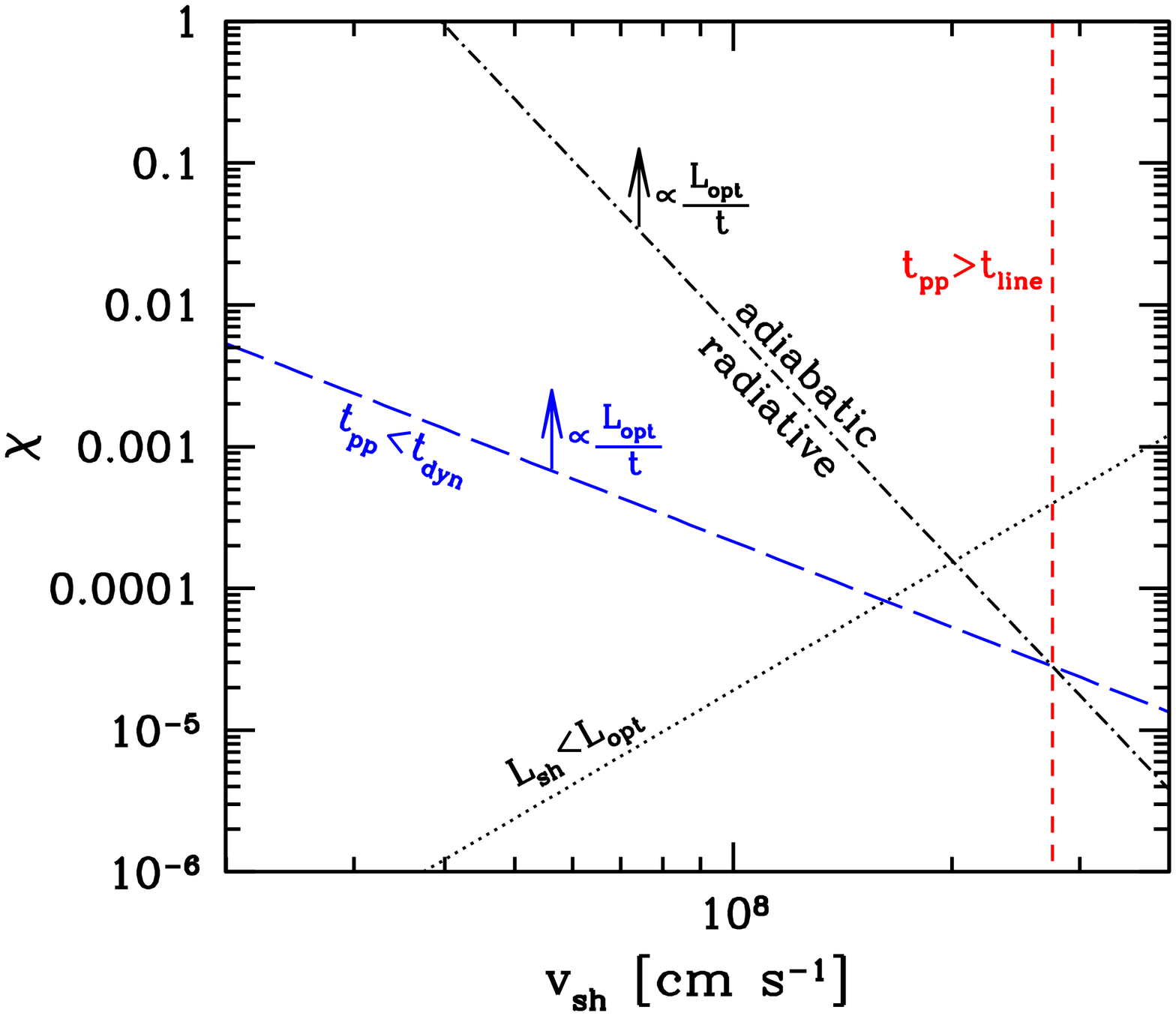}
  \vspace*{-0.6cm}
  \end{tabular}
  \end{center}
  \caption{Different cooling regimes of non-thermal leptons (left panel)
  and hadrons (right panel) in the parameter space of shock velocity $\vsh$ and $\chi$,
  the dimensionless ratio of radiation to matter energy density (Equation~(\ref{eq:chi}), calculated in the immediate downstream of the shock).
  The contours are computed for $\Lrad = 10^{38}$~erg~s$^{-1}$, $t = 1$~week, and lepton Lorentz factor $\gamma=10^3$,
  and assuming a shock radius $R = \vsh t$; the scalings of the contours with $\Lrad$, $t$ and $\gamma$ are shown on the plots.
  The black solid line separates regions where the radiative cooling of non-thermal electrons is dominated by bremsstrahlung or IC, respectively.
  The black dash-dotted line delineates regions where the thermal cooling/compression time is shorter/longer that the dynamical time (radiative vs.~adiabatic shock regime).
  Below the blue long-dashed line the radiative cooling time (leptons) and nuclear collision time (hadrons) is shorter than the dynamical time.  Below the red dashed line $\tcl < \min(\tcbr,\tcIC)$ (leptons) and $\tcl < \tcpp$ (hadrons), i.e. compression is faster than non-thermal losses.  The black dotted line (Equation \ref{eq:chi_min}) denotes where the total shock-generated luminosity equals the assumed optical luminosity; the region below the line is unphysical.
  }
  \label{Fig:cool}
\end{figure*}

It is instructive to compare the thermal cooling time behind the shock to the radiative and Coulomb loss times of the relativistic electrons/pairs.
For bremsstrahlung we find from Equations (\ref{eq:cool:l}) and (\ref{eq:gdotbr})
\begin{align}
\tcbr = 2.6 \times 10^5 \, \gamma_3^{-0.2} n_{9}^{-1} \,\, \mbox{s}, \quad \quad
\frac{\tcbr}{\tcl} = 33\, \gamma^{-0.2}_3 \, v_{{\rm sh}, 8}^{-3.4},
\end{align}
where $\gamma_3 = \gamma/10^3$, and $n$ is the upstream density.
Similarly, for Coulomb cooling one obtains (for $\gamma\gg 1$)
\begin{align}
\tcCoul = 3.3\times 10^5 \, \gamma_3 \, n_{9}^{-1} \, \left(\frac{\ln{\Lambda}}{25}\right)^{-1} \,\, \mbox{s}, \nonumber \\
\frac{\tcCoul}{\tcl} = 43 \, \gamma_3 \, v_{{\rm sh}, 8}^{-3.4} \,\left(\frac{\ln{\Lambda}}{25}\right)^{-1}.
\label{eq:tCrel}
\end{align}
Finally, for IC we find
\begin{align}
  \tcIC= \left\{\begin{array}{ll}
         1.2\times 10^6 \,  L_{38}^{-1}  \, R_{14}^2  \, \gamma_3^{-1} \, \mbox{s},  		& L_{\rm opt}>\Lsh \vspace{1mm}\\
         2.0\times 10^6\, n_{9}^{-1} v_{{\rm sh},8}^{-3} \, \gamma_3^{-1}  \, \mbox{s}, 		& L_{\rm opt}\approx \Lsh
                                       \end{array}
                               \right.
\end{align}
and
\begin{align}
&\frac{\tcIC}{\tcl} = \frac{4.7}{\chi\gamma \, v_{{\rm sh}, 8}^{3.4}}	\nonumber \\
&= 
\left\{ \begin{array}{ll}
	150 \, L_{38}^{-1}  \, R_{14}^2 \, n_{9} \, v_{{\rm sh}, 8}^{-3.4} \, \gamma_3^{-1},		& L_{\rm opt}>\Lsh \vspace{1mm}\\
	250 \, v_{{\rm sh}, 8}^{-6.4}  \, \gamma_3^{-1},						& L_{\rm opt}\approx \Lsh.
	\end{array}
	\right.
	\label{eq:tICrel}
\end{align}

For shock velocities $v_{\rm sh} \lesssim 2\times 10^8$~cm~s$^{-1}$ there exists a range of $\gamma$ over which particles are unable to cool before the plasma has strongly compressed.
In this regime, three different regions can be identified in electron energy space;
using our fiducial parameter set as an example, one finds:
(1) below $\gamma$ of a few tens, electrons rapidly share their energy with the thermal population via Coulomb interactions,
(2) above $\gamma\sim 10^5$ the electrons lose most of their energy via IC before the plasma has time to significantly compress, and
(3) in the intermediate range the electrons
undergo significant compression before cooling, and gain a moderate amount of additional energy from adiabatic heating.
This intermediate range of $\gamma$ is broader for slower (more radiative) shocks.  Conversely, in faster shocks the non-thermal particles cool faster than the thermal plasma can cool, such that the adiabatic heating described above does not arise (note that $\tcbr/\tcl$ is independent of density).

Figure \ref{Fig:cool} summarizes the different cooling regimes in the $\vsh-\chi$ parameter space.
The high observed luminosities of the {\it Fermi}/LAT emission from novae constrain the allowed region to reside
not too far above the dotted line, which denotes where the total shock power is comparable to the optical luminosity (\citealt{Metzger+15}).
In this region, the shock is radiative for velocities $\vsh\lesssim 2\times 10^8$~cm~s$^{-1}$
and thermal line cooling/compression is faster than either IC or bremsstrahlung cooling (red dashed line).
In the radiative shock regime, IC losses are at most comparable to bremsstrahlung losses for the range of electron energies $\gamma\lesssim 10^3$
that are later shown to be responsible for most of the hard X-ray emission.
Note that the relative dominance of bremsstrahlung losses over IC losses is further enhanced as the downstream plasma compresses.

\subsection{Non-thermal processes: hadronic scenario}

Protons which undergo diffusive shock acceleration are injected into the downstream with a distribution
$d\Np/d(\gp\bp) \propto (\gp\bp)^{-\qp}$, where $\qp = 2-2.5$ (\citealt{Blandford&Ostriker78}; \citealt{Caprioli&Spitkovsky14}),
which places most of the non-thermal energy into relativistic protons ($\Ep \gtrsim 1$~GeV).
In dense media they subsequently cool via hadronic collisions with thermal ions, on a timescale
\begin{align}
\tcpp \approx \frac{1}{c\spp \nds} = 2.5 \times 10^5 \, n_{9}^{-1} \,\, \mbox{s},
\end{align}
where $\spp \approx \sigmaT/20$, 
and we have again used $\nds = 4n$.  For $\vsh < 3\times 10^{8}$~cm~s$^{-1}$, downstream compression due to thermal cooling occurs faster than hadronic losses,
\begin{align}
\frac{\tcpp}{\tcl} = 32\, v_{{\rm sh}, 8}^{-3.4}.
\label{eq:tpprel}
\end{align}
The cooling regimes in the $\vsh - \chi$ parameter space are shown in Figure \ref{Fig:cool} (right panel).
Note that if the shock is radiative, the thermal cooling/compression behind the shock is always faster than the losses due to nuclear collisions.
The value of $\tcpp\propto n^{-1}$
decreases as the post-shock gas compresses,
leading to more efficient hadronic losses and extending the parameter range
over which the energy of the accelerated protons can be efficiently tapped.

Mildly relativistic protons ($\Ep \approx 1$~GeV) lose comparable fractions of their energy via elastic and inelastic collisions;
the elastic fraction decreases at higher energies.
The inelastic collisions produce both
neutral and charged pions ($\pi_{0}$, $\pi_{\pm}$),
which ultimately decay into GeV gamma-rays, $\epm$-pairs, and neutrinos ($\nue$, $\numu$, $\anue$, $\anumu$).
The spectra of the injected gamma-rays and pairs roughly mimic the slope of the primary protons (\citealt{Kamae+06});
for example, with $\qp=2$, their distribution is flat in energy per logarithmic interval of $\nu$ ($\gamma$).
Both the injected photon and electron spectra have a low-energy turnover at $\sim 100$~MeV.

The relativistic pairs 
cool via bremsstrahlung and IC emission (Section \ref{sec:nth:lep}), generating both X-ray and gamma-ray radiation.
Compared to the leptonic model,
the fraction of the total non-thermal energy emerging as hard X-rays is lower, for two main reasons:
(1) the fraction of energy of the injected $\epm$ pairs is only $\sim 10-20$~\% of the total energy dissipated via hadronic collisions, and
(2) the pair injection spectrum has a turnover at $\gamma\sim 200$.
The electrons that radiate in the hard X-ray band have Lorentz factors of $\gamma \approx 1 - 1000$;
in leptonic models the electrons are injected with comparable power throughout this range,
while only $\gamma \sim 100 - 1000$ leptons from $\pi_{\pm}$ decay can 
make an appreciable contribution in the hadronic case.

\section{Structure of the cooling layer}

\label{sec:gasdyn}

Consider the thermodynamic evolution of plasma in the downstream of a non-relativistic shock.
The gas pressure in the immediate downstream is dominated by thermal plasma.
In addition,
a fraction $\epsnth$, $\epsp$ and $\epsB$ of the energy is deposited
into non-thermal electrons, baryons, and the magnetic field, respectively 
(over scales much smaller than the post-shock cooling length).
The downstream plasma cools via radiation, both thermal and non-thermal, and compresses.
Assuming the radiative cooling timescale is shorter than the expansion time,
the total pressure in the cooling layer is approximately constant.

Consider a small (Lagrangian) volume element of plasma as it propagates into the downstream.
The first law of thermodynamics for this fluid element,
$\d (uV) = -p \d V + V \d Q_{\rm rad}$,
can be rewritten as
\begin{align}
\d u = h \, \d \ln{n} + \d Q_{\rm rad}.
\label{eq:du}
\end{align}
Here $u$ and $h=u+p$ are the total energy density and enthalpy, respectively, $n$ is the density, and $\d Q_{\rm rad}$ is
the total energy loss from the element (via radiation and neutrinos) per unit volume.

The assumption of constant downstream pressure imposes a constraint
\begin{align}
\d p = \d\pth + \d\pnth + \d\pprot + \d\pb = 0,
\label{eq:dp}
\end{align}
where $\pth$, $\pprot$, $\pnth$ and $\pb$ are partial pressures of the thermal plasma, non-thermal (accelerated) leptons, non-thermal protons, 
and the magnetic field, respectively.

Equations (\ref{eq:du}) and (\ref{eq:dp}) must be complemented by an equation of state (EOS) for each component of the plasma,
i.e. $h_i = \alpha_i u_i = \alpha_i p_i/(\alpha_i-1)$, where $\alpha_i$ is the adiabatic index.
For the thermal component, $\ath=5/3$.
The adabatic index for the magnetic field depends on the field configuration;
 $\ab=4/3$ for a tangled field with a coherence length much smaller than the cooling length,
while $\ab=2$ for an ordered field perpendicular to the direction of compression\footnote{Strong one-dimensional compression enhances the perpendicular component of the magnetic field;
consequently the adiabatic index of an initially random field evolves towards $\ab=2$,
unless the random component decays faster than the compression time.}  (i.e. shock normal).
The EOS of the non-thermal components of the plasma do not admit the above simple form,
as they contain both non-relativistic and relativistic particles, and hence must be found explicitly at each timestep from the solution of the non-thermal evolution (see below). 

Equations (\ref{eq:du}) and (\ref{eq:dp}) can be transformed to read
\begin{align}
\d\ln{n} = \frac{(\ath - 1)(\d\unth + \d\uprot -\d Q_{\rm rad})  - (\d\pnth + \d \pprot) }{(\ath-1) h + (\ab - \ath) h_B },
\label{eq:dn}
\end{align}
where we have used the EOS described above for the thermal plasma and the magnetic field, as well as $\d\ub = \hb \, \d\ln{n}$ for the adiabatically evolving $B$-field.
Coupled with equations for the energy spectra of non-thermal electrons and protons (which provide $\d\unth$, $\d\pnth$, $\d\uprot$, $\d\pprot$),
Equation (\ref{eq:dn}) determines the downstream evolution of the shocked plasma.
Here $\d Q_{\rm rad}$ accounts for radiative (and neutrino) losses of both thermal and non-thermal plasma, but does not explicitly involve coupling
(e.g. Coulomb) between the themal and non-thermal particles.
Once $\d n$ is known, the updated magnetic pressure is found from $\pb\propto n^{\ab}$; 
condition (\ref{eq:dp}) then yields the new $\pth$.  We neglect any reconnection or decay of the magnetic field, e.g. as could occur due to ambipolar diffusion once the temperature cools to $\sim 10^{4}$ K and the gas becomes neutral.

\begin{figure*} 
  \begin{center}
  \begin{tabular}{cc}
  \includegraphics[trim = 0cm 0cm 0cm 0cm, width=0.45\textwidth]{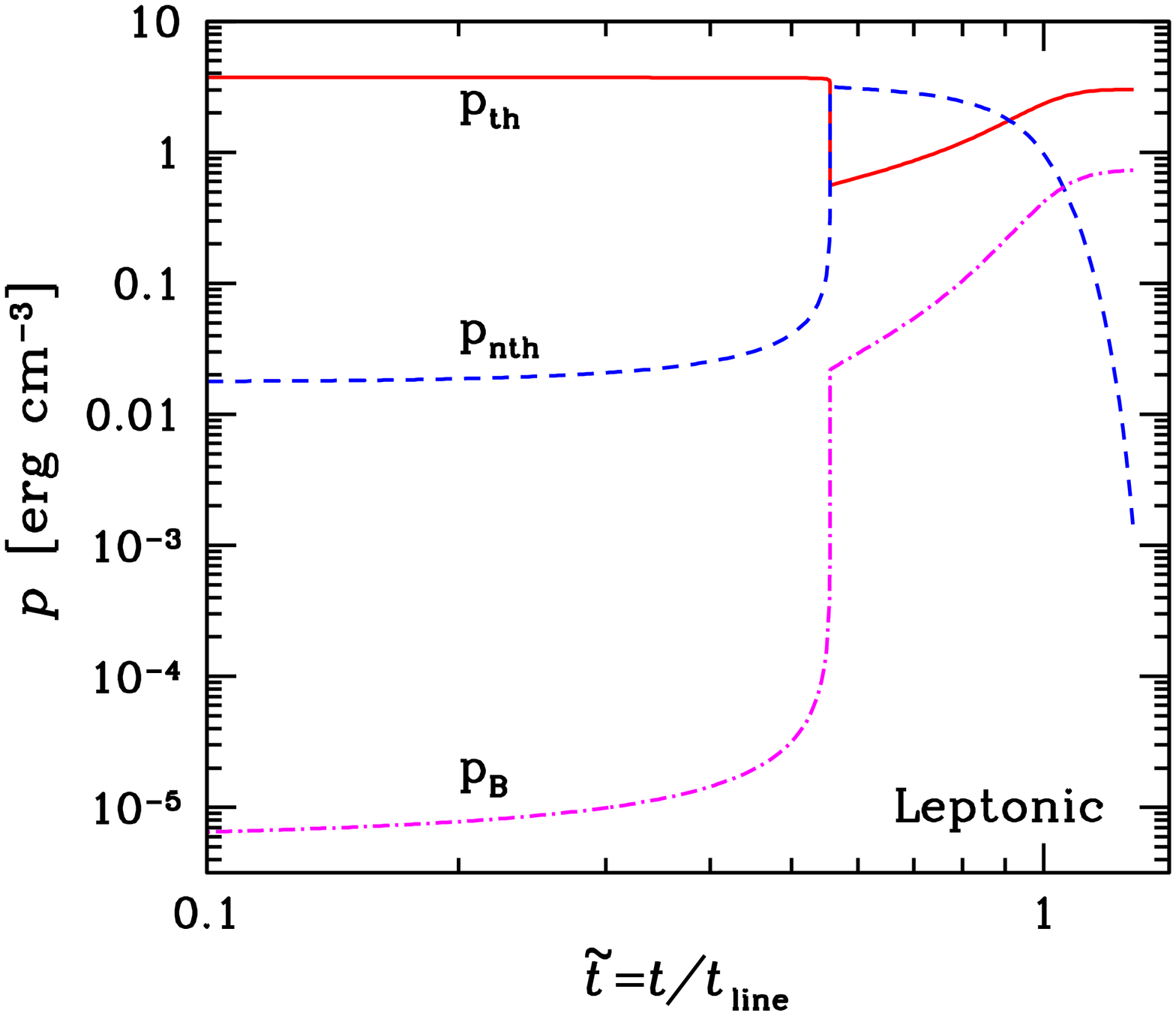} &
  \includegraphics[trim = 0cm 0cm 0cm 0cm, width=0.45\textwidth]{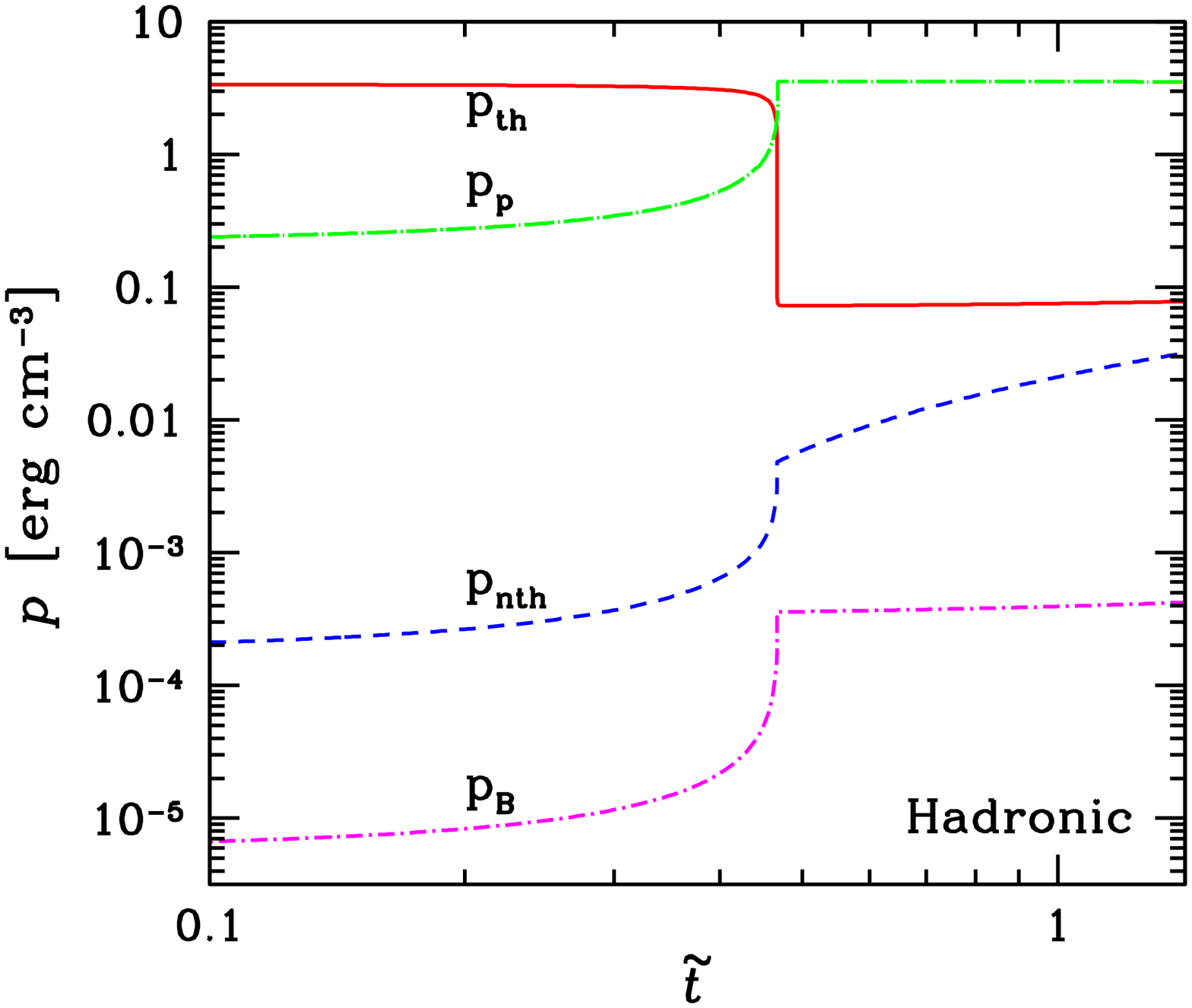}
  \end{tabular}
  \end{center}
    \vspace*{-0.4cm}
  \caption{Evolution of downstream pressure.
  Parameters:
  pre-shock density $n = 3\times 10^8$~cm$^{-3}$,
  shock velocity $v_{\rm sh}= 10^{8}$~cm~s$^{-1}$,
  $\chi=10^{-4}$ (which corresponds to $L=10^{38}$~erg~s$^{-1}$, $R=10^{14}$~cm).
  Left panel: leptonic case. Nonthermal injection fraction $\epsnth=0.01$, magnetization $\epsB=10^{-6}$.
  Nonthermal electrons are injected at the shock with a distribution
  $dN/d(\gamma\beta)\equiv \Qe(\gamma\beta)\propto (\gamma\beta)^{-q}$,
  where $q=2$ and the distribution extends to $\gamma_{\rm max}=10^5$.
  Right panel: hadronic case. nonthermal injection fraction $\epsp=0.1$, magnetization $\epsB=10^{-6}$.
  Nonthermal protons are injected with a distribution $d\Np/d(\gp\bp)\propto (\gp\bp)^{-\qp}$,
  where $\qp=2$ and the distribution extends to $\gpmax=10^3$.
  A weak nonthermal electron distribution with $\epsnth=10^{-4}$ is also injected, with the same distribution as in the leptonic case.
  Adiabatic index for the magnetic field $\ab=2$ in both panels.
   The Lagrangian time of a fluid element on the $x$-axis
  corresponds to coordinate $z=\int^t v(t^{\prime}) \, dt^{\prime}$ (assuming $t\ll \texp$).}
  \label{Fig:p}
\end{figure*}

\begin{figure*} 
  \begin{center}
  \begin{tabular}{cc}
  \includegraphics[trim = 0cm 0cm 0cm 0cm, width=0.45\textwidth]{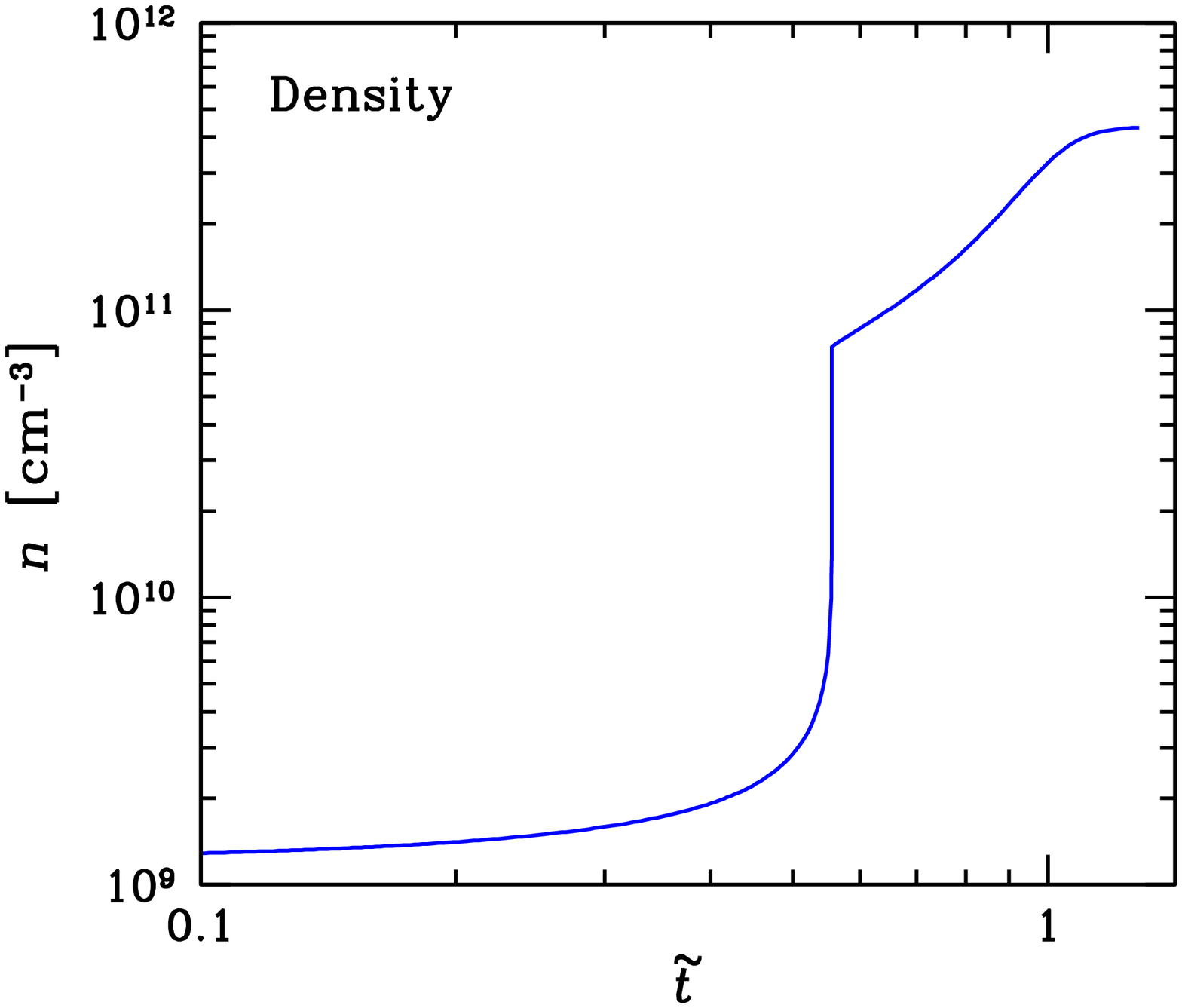}	&
  \includegraphics[trim = 0cm 0cm 0cm 0cm, width=0.45\textwidth]{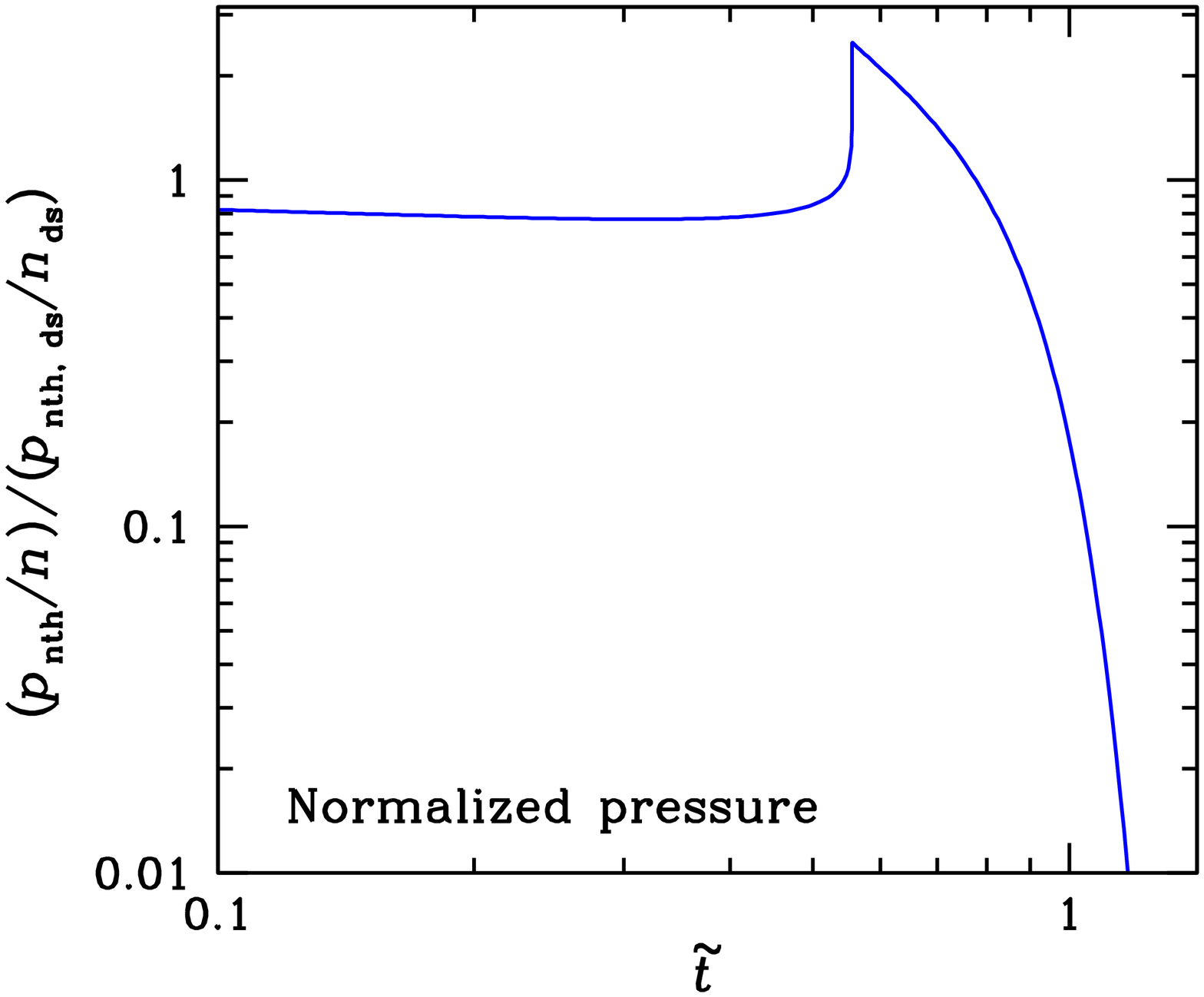}
  \end{tabular}
  \end{center}
  \vspace*{-0.4cm}
  \caption{
  Left panel: evolution of the downstream density.
  Right panel: nonthermal pressure normalized to unit density, relative to the postshock value.
  Parameters are the same as in the left panel of Figure \ref{Fig:p}.
}
  \label{Fig:n}
\end{figure*}

Figures \ref{Fig:p} (left panel) and \ref{Fig:n} show the evolution of the pressure and density in the cooling layer in the leptonic model ($\pprot = 0$).
Immediately behind the shock, the bulk of the dissipated energy is stored in thermal plasma;
its (line-)cooling therefore determines the downstream evolution
over the first thermal cooling time.  
The plasma initially compresses as $n\propto T^{-1}$ at $\pth\approx$~constant; the decreasing temperature and increasing density
speed up the cooling  ($\tcl\propto T^{1.7} n^{-1}$), which leads to runaway loss of thermal pressure.

Figure \ref{Fig:p} (left panel) shows that
most relativistic electrons
are unable to cool over the compression timescale $\tcl$ and thus retain most of their energy/pressure.
As the plasma compresses, both the nonthermal and magnetic pressure increase at the expense of the thermal energy.
Depending on the initial value of $\epsnth/\epsB$, eventually either $\pnth$ or $\pb$ comes to dominate,
thus controlling the evolution further downstream.  The particular model shown in Fig.~\ref{Fig:p} has $\pnth > \pb$ at the time the thermal plasma has cooled.
In this case, the plasma continues to compress on the nonthermal cooling timescale (Figure \ref{Fig:n}, left panel).
Eventually the non-thermal pressure is also lost due to the cooling of relativistic particles and 
further compression is only halted once magnetic and/or thermal pressure (from the cooled dense plasma at $T\approx 10^4$~K) comes to dominate.

In the hadronic scenario (Figure \ref{Fig:p}, right panel)
the downstream plasma initially compresses until the pressure becomes dominated by relativistic protons;
further compression after the loss of thermal support is governed by their cooling via nuclear collisions.

Consider, as a point of contrast, a case in which the nonthermal particles are fast-cooling even when the thermal plasma is not, as can occur in high velocity, adiabatic shocks.
In such a case the nonthermal pressure is lost before significant compression has taken place and thus it plays no role in the evolution of the downstream plasma.
In turn, the nonthermal emission is unaffected by the thermal evolution of the post-shock gas.

\section{Non-thermal evolution}

\label{sec:nthevo}

The distribution functions $N(\gamma)$ of non-thermal leptons and protons evolve with distance $z$ downstream of the shock according to 
\begin{align}
\frac{\partial N(\gamma)}{\partial t} + \frac{\partial}{\partial z} [v N(\gamma) ]+ \frac{\partial}{\partial \gamma} [\dot{\gamma} N(\gamma)] = \Qinj,
\label{eq:Nnth0}
\end{align}
where $v$ is the downstream velocity, and $\dot{\gamma}$ is the energy loss/gain rate
due to all interactions, in units of the particle rest energy per second.  Here $N(\gamma)$ is defined such that $n_{\rm nth}$ or $n_{\rm p} =\int N(\gamma) d\gamma$.
For leptons, $\dot{\gamma}$ accounts for IC, bremsstrahlung and Coulomb losses,
as well as adiabatic heating/cooling\footnote{Here all processes are treated as continuous;
this assumption breaks down for IC emission in the Klein-Nishina regime, which becomes relevant at $\gamma\gtrsim 10^4$.
However, the {\it average} energy loss rate over several scatterings is still accurate, provided that $\dot{\gamma}$ appropriately accounts for KN suppression.}.
For protons, $\dot{\gamma}$ represents losses due to nuclear collisions, Coulomb scattering, as well as heating by adiabatic compression.
The right hand side accounts for injection of pairs due to hadronic collisions; $\Qinj=0$ in the proton equation.

Equation (\ref{eq:Nnth0}) can be cast in a more convenient form by expanding the second term and using the continuity equation
$\partial n/\partial t + \partial (v n)/\partial z = 0$, 
\begin{align}
\frac{d}{dt} \left[ \frac{N(\gamma)}{n} \right] = -\frac{\partial}{\partial\gamma}
\left[ \frac{ \dot{\gamma} N(\gamma)}{n} \right] + \frac{\Qinj}{n},
\label{eq:Nnth}
\end{align}
where the time derivative is now taken along the path of the fluid element as it propagates into the downstream. 
Here $n$ represents the inverse of a the comoving volume element (such that $nV = \mbox{constant}$), or equivalently,
the density of any type of particle whose total number is conserved (e.g. baryons).

The rate of adiabatic heating/cooling of a particle of energy $\gamma$, as determined by considering the first law of thermodynamics for a monoenergetic particle distribution, is given by
\begin{align}
\dot{\gamma}_{\rm adiab} = \frac{1}{3} \gamma\beta^2 \, \frac{d\ln{n}}{dt},
\end{align}
where $\beta =v/c$.

Given $N(\gamma)$, the non-thermal pressure and (kinetic) energy densities $\pnth$, $\unth$, $\pprot$, $\uprot$ are found from
\begin{align}
p &= \frac{1}{3} mc^2 \int_1^\infty N(\gamma) \, \gamma\beta^2 \, d\gamma, \nonumber \\
u &= mc^2 \int_1^\infty N(\gamma) \, (\gamma-1) \, d\gamma,
\end{align}
where $m$ is the particle mass.

Equations (\ref{eq:dn}) and (\ref{eq:Nnth}) are coupled via $n$, $\pnth$, $\unth$, $\pprot$, $\uprot$ and are solved iteratively at each step.
In the radiative regime the downstream structure at any given radius is given by
$n(z)$ and $N(\gamma,z)$, where $z=\int^t v(t^{\prime}) \, dt^{\prime}$ and
 $v = \vds \nds/n$.
Here $\vds$ is the velocity relative to the shock front and $\nds$ is the density, both measured
in the immediate downstream.
The solutions $N(\gamma,z)$ for both electrons and protons
determine the emissivities due to different processes
throughout the cooling layer; the emerging spectra are obtained by integrating the emissivities over $dz$.

\begin{figure} 
  \begin{center}
  \begin{tabular}{c}
    \includegraphics[width=0.45\textwidth]{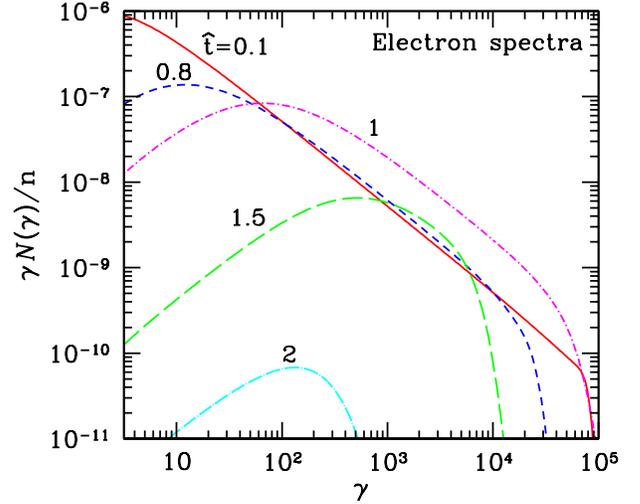}
  \vspace*{-0.4cm}
  \end{tabular}
  \end{center}
  \caption{Electron distribution at different Lagrangian times behind the shock, normalized to the overall density.
  The labels show time normalized to the time of fastest compression ($\hat{t} \approx 2 \, t/\tcl$).
  The red solid line approximately corresponds to the initial distribution at the shock.
  The magenta (dash-dotted) line shows the distribution immediately after the thermal pressure is lost.
   Parameters: shock velocity $v_{\rm sh}= 10^{8}$~cm~s$^{-1}$,
  $\chi=10^{-3}$, nonthermal injection fraction $\epsnth=0.01$, magnetization $\epsB=10^{-6}$, pre-shock density $n = 3\times 10^7$~cm$^{-3}$.}
  \label{Fig:el}
\end{figure}

Figure \ref{Fig:el} shows snapshots of the cooling nonthermal electron distribution at different times/distances behind the shock ($\pprot = 0$).
The blue dashed line shows the distribution at a time $\tilde{t}=0.8$ just before the thermal pressure is lost, i.e. prior to the fastest compression.
The deficiency of electrons at low and high energies arises from their fast cooling relative to the compression/thermal cooling time, due to Coulomb and IC losses, respectively (Equations~\ref{eq:tCrel} and \ref{eq:tICrel}).
Rapid compression at $\tilde{t}=1$ shifts the entire distribution towards higher energies,
by a factor $\propto n^{1/3} \sim 2$ (the apparent shift is larger at the low energy end due to significant Coulomb losses between $\tilde{t}=0.8$ and $1$).
At later times the re-energized distribution rapidly cools owing to the increased rates $\dgbr\propto\dgC\propto n$;
the cooling is slowest at $\gamma\approx \gamma_{\star}$ where $\dgbr\approx\dgC$ (Equation \ref{eq:gstar}).

\section{High-energy radiation from the cooling layer}

\label{sec:emission}

Relativistic leptons, either accelerated directly at the shock or produced by $\pi_{\pm}$ decay, give rise to emission extending from the X-ray to gamma-ray band.  In the hadronic scenario, however, the dominant source of $>100$~MeV gamma-rays is the decay of neutral pions ($\pi_0$).  The rates of these processes must be calculated self-consistently with the evolution of the post-shock plasma. 

 In broad terms, one can identify two main regions of the cooling layer which are 
relevant for the spectral formation:
(1) the first thermal cooling length, where
the plasma density and partial pressures
are approximately equal to their immediate post-shock values, and
(2) the high-density region further downstream.
As noted above, compression deposits additional energy into the non-thermal population;
this is seen as a jump at $t\approx 10^4$~s on Figure \ref{Fig:n} (right panel), where we show the nonthermal pressure normalized to unit density (which approximately characterizes the energy per particle).
Compression by a factor $\sim 10$ results in approximate energy gain of $\propto n^{1/3} \sim 2$ per particle.  The increased density also enhances the bremsstrahlung and Coulomb cooling rates relative to IC
($\dgbr/\dgIC\propto \chi^{-1} \propto n$).  Consequently, the additional energy deposited into the nonthermal leptons via adiabatic heating mainly enhances the bremsstrahlung spectral component.

\subsection{Gamma-ray spectrum}

If the downstream magnetization is relatively weak, electrons/pairs of energy $\gamma \gtrsim 10^3$ cool primarily by bremsstrahlung or IC radiation.
The characteristic energies of the emitted photons are $E \approx \gamma \me c^2 \approx 5 \, \gamma_4$~GeV (bremsstrahlung) and
$E \approx (4/3)\gamma^2 \Eopt  \approx 0.3 \gamma^2_4 (\Eopt/2\,\mbox{eV})$~GeV (IC),
where $\Eopt$ is the average energy of optical seed photons, and $\gamma_4 \equiv \gamma/10^4$.

Thus, absent significant synchrotron losses, gamma-ray emission above a few
hundred MeV
serves as a calorimeter for the particle acceleration efficiency (\citealt{Metzger+15}).
For leptons injected with a distribution
$\Qe \equiv dN/d\gamma \propto \gamma^{-q}$,
the high-energy
spectrum in the fast cooling regime follows (defining $x\equiv h\nu/\me c^2$)
\begin{align}
&\nu F_{\nu} \propto \frac{dE_{\rm br}}{d\ln{x}} = x^{-q + 2},
\quad
&\frac{dE_{\rm IC}}{d\ln{x}} = x^{-(q - 2)/2 }
\end{align}
in the bremsstrahlung and IC-dominated cases, respectively.
The expected approximately flat injection spectrum ($q \approx 2$)
results in a similarly flat GeV spectrum regardless of the emission mechanism, 
in broad agreement with observations of novae \citep{Ackermann+14}.

In the hadronic scenario
the spectrum of $\pi_0$-decay gamma rays roughly mimics that of the accelerated protons, i.e.
$\nu F_{\nu} \propto \nu^{-\qp + 2}$;
the same is true for injected pairs from $\pi_{\pm}$ decay, for which $q \approx \qp$.
Thus a flat GeV spectrum is also expected in hadronic models.

One concludes that
the shape of the GeV spectrum alone is insufficient to distinguish between leptonic and hadronic models,
or between IC and bremsstrahlung origin of the GeV radiation in the former.

\subsection{X-ray spectrum}

\begin{figure*} 
  \begin{center}
  \begin{tabular}{cc}
  \includegraphics[trim = 0cm 0cm 0cm 0cm, width=0.45\textwidth]{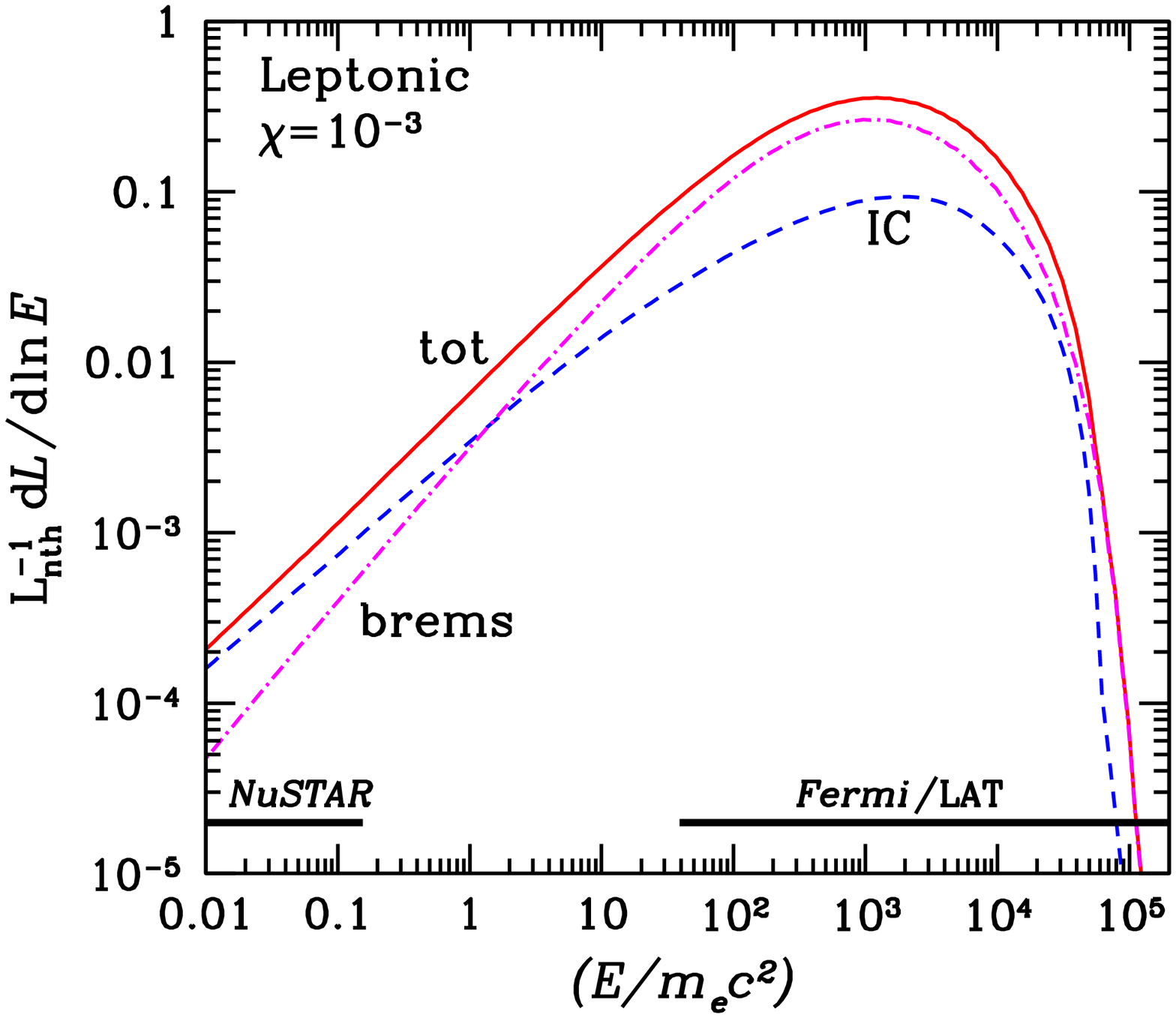}	&
  \includegraphics[trim = 0cm 0cm 0cm 0cm, width=0.45\textwidth]{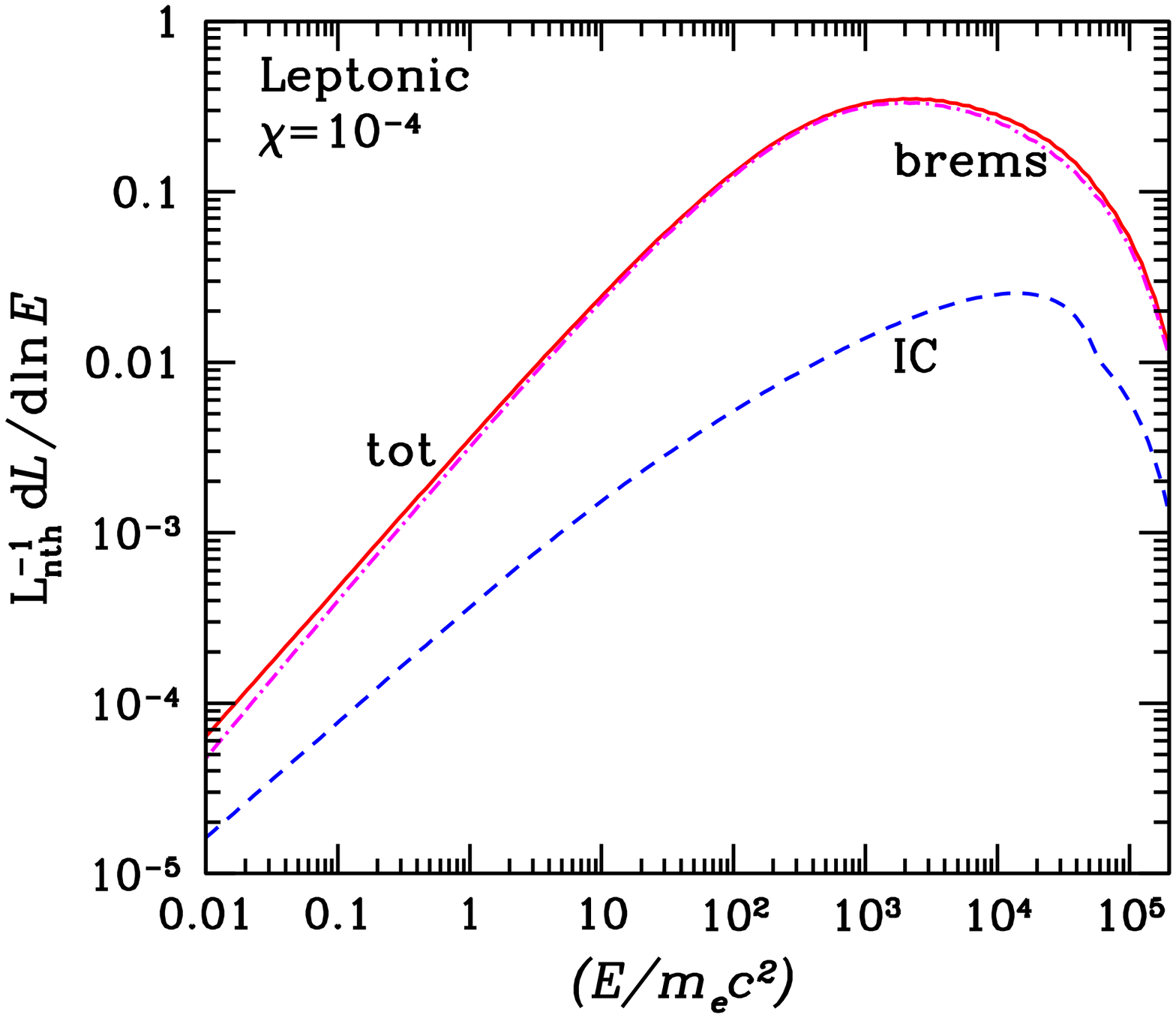}
  \end{tabular}
  \end{center}
  \vspace*{-0.4cm}
  \caption{Shock spectra in the leptonic model, normalized to the total energy injected into nonthermal electrons.
  The inverse Compton spectrum is shown in blue (dashed line), bremsstrahlung in magenta (dash-dotted line), and
  total spectrum in red (solid line).
   Parameters: shock velocity $v_{\rm sh}= 10^{8}$~cm~s$^{-1}$,
  nonthermal injection fraction $\epsnth=0.01$, magnetization $\epsB=10^{-6}$.
  Injected electron distribution
  $\Qe(\gamma\beta)\propto (\gamma\beta)^{-q}$,
  where $q=2$ and the distribution extends to $\gamma_{\rm max}=10^5$.
   Left panel:
   $\chi=10^{-3}$ (corresponds to pre-shock density $n = 3\times 10^7$~cm$^{-3}$, if $\Lrad=10^{38}$~erg~s$^{-1}$ and $R=10^{14}$~cm),
   right panel:
   $\chi=10^{-4}$.
   }
  \label{Fig:sp}
\end{figure*}

\begin{figure} 
  \begin{center}
  \begin{tabular}{c}
  \includegraphics[trim = 0cm 0cm 0cm 0cm, width=0.45\textwidth]{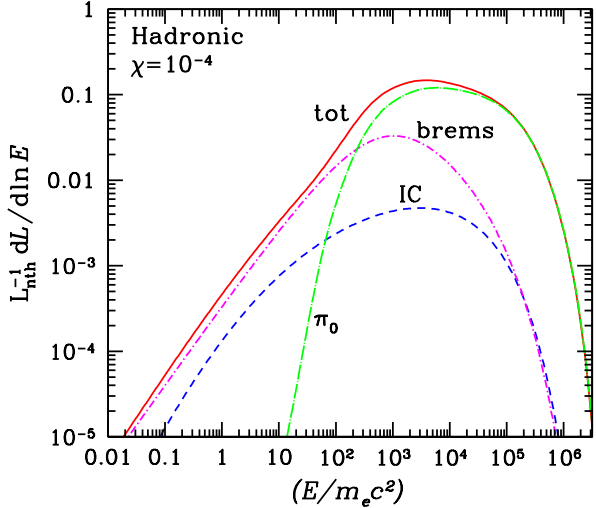}
  \end{tabular}
  \end{center}
  \vspace*{-0.4cm}
  \caption{Shock spectra in the hadronic model, normalized to the total energy injected into nonthermal protons.
  Blue dashed line: inverse Compton, magenta dash-dotted line: bremsstrahlung,
  green long dash-dotted line: $\pi_0$ decay,
  red solid line: total spectrum.
   Parameters: shock velocity $v_{\rm sh}= 10^{8}$~cm~s$^{-1}$,
   $\chi=10^{-4}$ (corresponds to pre-shock density $n = 3\times 10^8$~cm$^{-3}$, if $\Lrad=10^{38}$~erg~s$^{-1}$ and $R=10^{14}$~cm),
  nonthermal injection fraction $\epsp=0.1$, magnetization $\epsB=10^{-6}$.
  Injected proton distribution
  $\Qp(\gp\bp)\propto (\gp\bp)^{-\qp}$,
  where $\qp=2$ and the distribution extends to $\gpmax=10^3$.
  }
  \label{Fig:sp:hadr}
\end{figure}

Electrons/pairs of energy $\gamma \lesssim 10^3$ suffer significant Coulomb losses behind the shock,
which dramatically suppresses the hard X-ray emission.
Absent Coulomb losses, the relatively flat gamma-ray spectrum from $\Qe \propto \gamma^{-2}$ electrons
in leptonic models
would extend down into the X-ray band,
in which case the hard X-ray luminosity would rival that in the LAT bandpass.
In practice, this limit is attained only for unrealistically high values of $\chi \sim 1$, such that IC dominates over all other cooling mechanisms.
For more physical values of $\chi \ll 1$, Coulomb losses suppress the X-ray emission by several orders of magnitude.

Let us estimate the energy radiated in a given frequency band, such as hard X-rays,
by the rapidly cooling relativistic electrons behind the shock.
The total energy loss rate of a relativistic electron is
\begin{align}
\dg = \dgbr+\dgC+\dgIC
\label{eq:Q}
\end{align}
where the appropriate rates are given by Equations (\ref{eq:gdotbr}), (\ref{eq:gdotIC}) and (\ref{eq:gdotC}), respectively.

The bremsstrahlung emissivity at frequency $x\equiv h\nu/(\me c^2)\ll\gamma$ can be written as (e.g. \citealt{Haug97})
\begin{align}
\jbrep(x) \approx \frac{2}{\pi} c\sigmaT\afs \, n \,
\sum_i \frac{X_i Z_i^2}{A_i}
\int \ln\left(\frac{1.2\gamma^2}{x}\right) \, N(\gamma) \, d\gamma,
\label{eq:jbrep}
\end{align}
where hereafter we drop the sum over ion species in our estimates, assuming a hydrogen-dominated composition.

Similarly, using the delta-function approximation (as justified for a smooth electron distribution softer than $N(\gamma)\propto \gamma$), the IC emissivity can be written as
\begin{align}
\jIC(x) &\approx
\frac{4\sigmaT\urad}{3\me c} \int \delta\left( x - \frac{4}{3}\gamma^2 \xopt \right) \, \gamma^2 N(\gamma) \, d\gamma \nonumber \\
&=
\frac{1}{2}\, c\sigmaT\nrad
\gamma_0 N(\gamma_0),
\label{eq:jIC}
\end{align}
where $\gamma_0 \equiv (3x/4\xopt)^{1/2}$ and $\xopt \equiv \bar{E}_{\rm opt}/m_e c^{2}$ is the energy of the optical/UV seed photons.

The emissivity of a single electron of energy $\gamma$ is obtained by using $N(\gamma^{\prime})=\delta(\gamma^{\prime}-\gamma)$
in Equations (\ref{eq:jbrep}) and (\ref{eq:jIC});
the energy emitted by the electron at a given $x$
over its cooling history is then found by
taking the ratio of Equations (\ref{eq:jbrep}) (or (\ref{eq:jIC})) and (\ref{eq:Q}), 
and integrating over the energy $\gamma$ of the cooling electron%
\footnote{Even though Equation (\ref{eq:jIC}) is a very poor representation of the IC spectrum for monoenergetic electrons,
integration over the electron cooling history has the same effect as considering a broad electron spectrum,
and yields a sufficiently accurate result for our analytical estimates. Exact emissivities are used in the numerical calculations below.}.
For concreteness, we focus on electron energies $\gamma\lesssim 10^3$, for which Coulomb losses dominate the total cooling rate.
For bremsstrahlung emission, we obtain the spectrum of a single cooling electron
\begin{align}
&\left.
\frac{dE_{\rm br}}{d\ln{x}}\right|_{\rm 1\, el.} \simeq
\int_1^{\gamma} \frac{x\jbrep(x)}{\dgC} \, d\gamma 	\nonumber \\
&= \frac{4}{3\pi} \frac{\afs}{\ln\Lambda} \, x\int_1^{\gamma} \ln\left(\frac{1.2\gamma^2}{x} \right) \, d\gamma
\approx \frac{4}{3\pi} \frac{\afs}{\ln\Lambda} \, x\gamma \ln\left(\frac{1.2\gamma^2}{x} \right),
\label{eq:Ebr_e}
\end{align}
where in the last equality we have used the fact that the integral is dominated by contributions from high $\gamma$.
Now, considering emission from the entire injected electron distribution $\Qe(\gamma)$, the total emitted energy per frequency interval $\ln{x}$ is given by
\begin{align}
&\frac{dE_{\rm br}}{d\ln{x}} =
\left.
\int \frac{dE_{\rm br}}{d\ln{x}}\right|_{\rm 1\, el.} \, \Qe(\gamma)\, d\gamma \nonumber \\
&= 
 \frac{4}{3\pi} \frac{\afs}{\ln\Lambda} \, x \int_{\sim 1}^{\gmax} \ln\left(\frac{1.2\gamma^2}{x} \right) \, \Qe(\gamma)\,  \gamma\, d\gamma
 \label{eq:Ebr}
\end{align}
The upper boundary of the integral should be taken as the energy above which Coulomb collisions no longer dominate the cooling, i.e. $\gmax = \gamma_{\star}$ (Equation~\ref{eq:gstar}).
Unless the injected distribution is strongly inverted\footnote{For instance, $\Qe(\gamma)\propto \gamma^{-q}$, with $q\le 1$, as can be appropriate if the injected distribution has a low-energy cutoff,
e.g. as a result of $\pi^{\pm}$ decay.} the additional contributions from higher $\gamma$ (where cooling is dominated by either bremsstrahlung or IC)
can be shown to be at most comparable to (\ref{eq:Ebr}).
Equation (\ref{eq:Ebr}) is notably independent of shock parameters, other than $\Qe(\gamma)$.

For IC emission, a similar argument leads to
\begin{align}
&\left.
\frac{dE_{\rm IC}}{d\ln{x}}\right|_{\rm 1\, el.} = 
\int \frac{x\jIC(x)}{\dgC} \, d\gamma \nonumber \\
&= \frac{8}{9} \frac{\chi}{\ln\Lambda} \,
x\int \delta\left( x - \frac{4}{3}\gamma^2 \xopt \right) \gamma^2 \, d\gamma
\approx \frac{1}{3} \, \frac{\chi}{\xopt\ln\Lambda} \, x\gamma_0,
\label{eq:FIC_e}
\end{align}
where again $\gamma_0 \equiv (3x/4\xopt)^{1/2}$, but here $\chi = \urad/(\me c^2 n)$ is defined using the {\it local} density at the emission site (rather than the upstream density).
The IC spectrum from the entire electron population is thus given by
\begin{align}
\frac{dE_{\rm IC}}{d\ln{x}} = 
\frac{1}{3} \, \frac{\chi}{\xopt\ln\Lambda} \, x\gamma_0 \int_{\gamma_0} \Qe(\gamma) \, d\gamma,
\label{eq:EIC}
\end{align}
where the lower integration limit follows because only electrons  injected above $\gamma_0$ contribute to the flux at frequency $x$. 

Figure \ref{Fig:sp} shows the emission spectra from the cooling layer for two different upstream densities,
calculated assuming a logarithmically flat injected electron energy spectrum ($q=2$).
As expected from Equation (\ref{eq:Ebr}), the bremsstrahlung component
below the spectral peak is almost independent of density,
whereas the IC spectrum scales as $\chi\propto n^{-1}$ (Equation (\ref{eq:FIC_e})).
For both cases in Figure \ref{Fig:sp}, the thermal plasma cools faster than $\gamma \sim 10^3$ electrons, i.e.~$\tcl < \min(\tcIC,\tcbr)$;
the majority of the non-thermal emission therefore originates from the cooled and compressed layer,
where the density is $\sim 100$ times higher than its value near the shock.
The relevant value of $\chi$ one should use in the analytical estimate (\ref{eq:EIC}) is thus lower by the same factor, compared to its value near the shock (Equation \ref{eq:chichar}).

For an injection slope $q=2$, the bremsstrahlung X-ray emission receives approximately equal contributions
per logarithmic electron energy interval up to $\gamma \approx \gamma_{\star} \approx 10^3$;
the X-ray spectrum thus approximately follows $\nu F_{\nu}\propto \nu$ in both the IC- and bremsstrahlung-dominated cases.
The IC component is slightly softer, as the hard X-ray band is not far below the smooth spectral break that occurs when the emitting electrons
are no longer cooled by Coulomb collisions ($\gamma\sim 10^3$, which corresponds to IC photons of $\xopt\gamma^2 \approx$ a few MeV).

The arguments leading to Equations (\ref{eq:Ebr}) and (\ref{eq:EIC}) apply equally well for hadronic models,
except that the lepton injection $\Qe(\gamma)$ now occurs in a volume rather than at the shock.
The most significant difference is the lack of injected pairs below
$100$~MeV, which effectively introduces a lower integration limit of $\gmin \approx 200$ in Equation (\ref{eq:Ebr}).
The narrower integration range $\gamma\in [\gmin,\gamma_{\star}]$ instead of $\gamma\in [1,\gamma_{\star}]$ 
lowers the hard X-ray flux by a logarithmic factor of a few in the hadronic case (assuming $\Qe(\gamma)\propto \gamma^{-2}$).

\section{X-ray to gamma-ray luminosity ratio}

\label{sec:Xgratio}

\begin{figure*} 
  \begin{center}
  \begin{tabular}{cc}
  \includegraphics[trim = 0cm 0cm 0cm 0cm, width=0.45\textwidth]{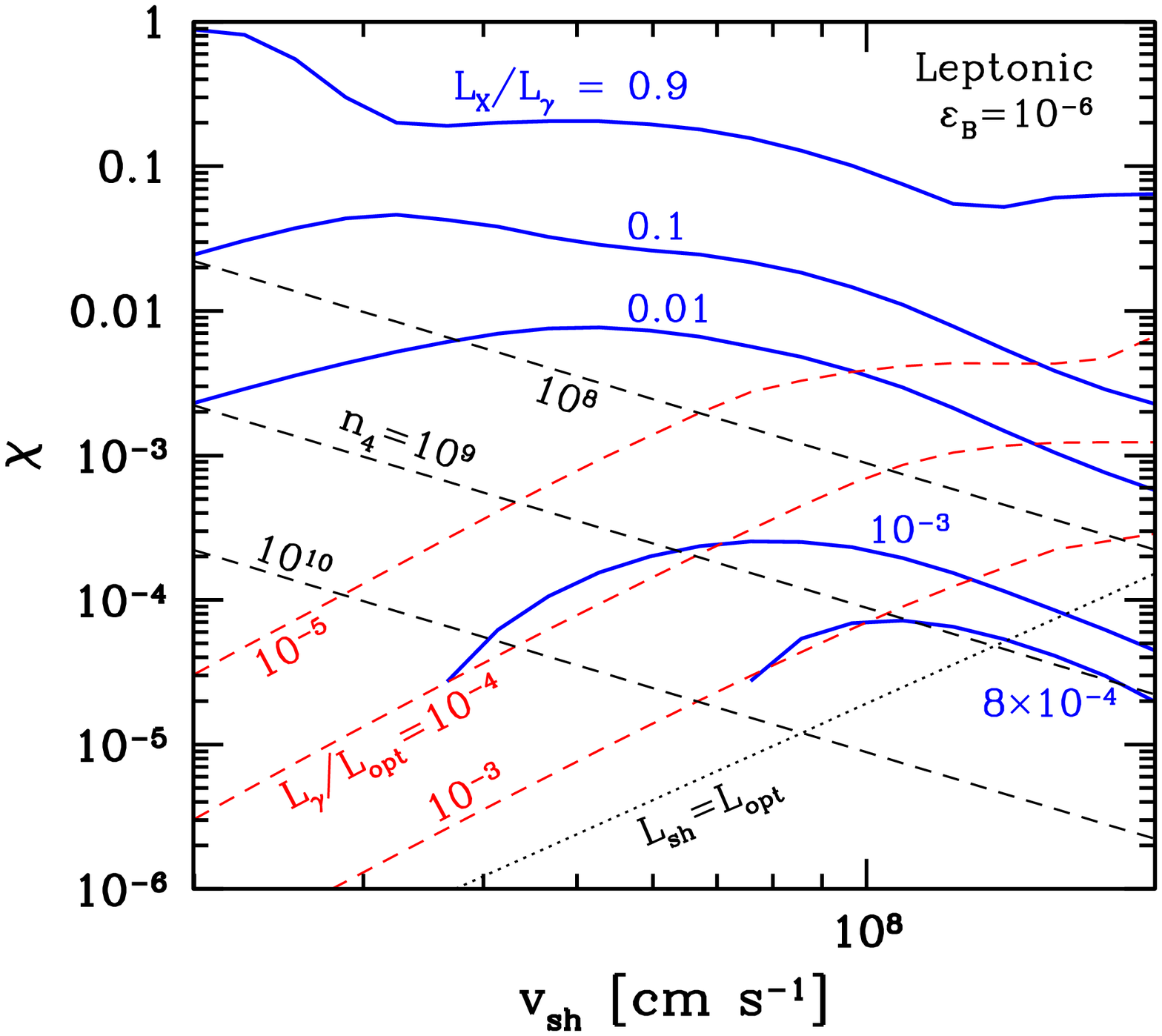}	&
  \includegraphics[trim = 0cm 0cm 0cm 0cm, width=0.45\textwidth]{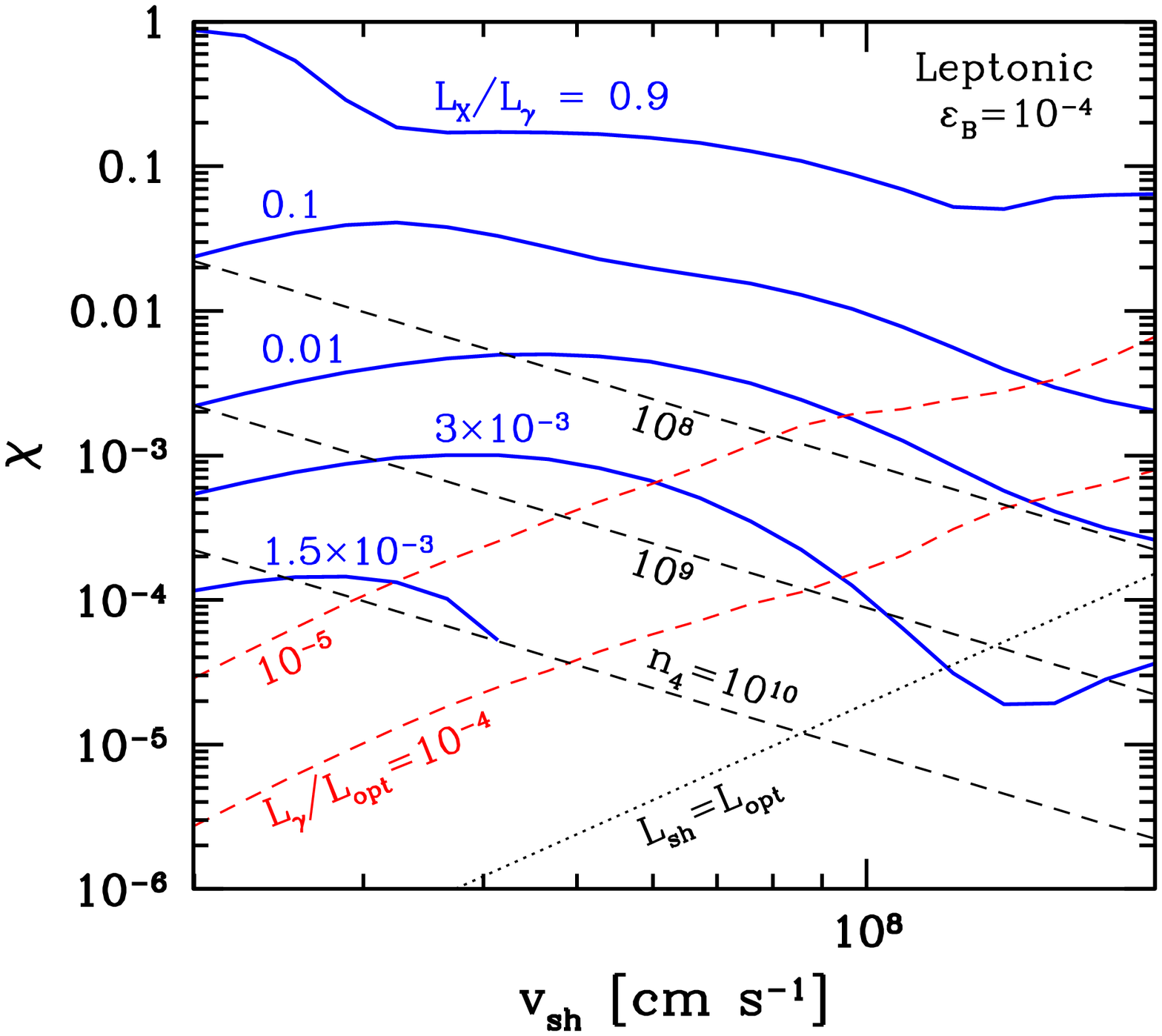}	
  \end{tabular}
  \end{center}
  \vspace*{-0.4cm}
  \caption{
  Isocontours of relative fluxes at $30$~keV and $1$~GeV, $\nu F_{\nu, 30 \mathrm{keV}}/\nu F_{\nu, 1 \mathrm{GeV}}$ in leptonic models (blue solid lines), in the parameter space of shock velocity $\vsh$ and the compactness to Thomson opacity ratio $\chi=\urad/(m_e c^2 n)$ (Equation (\ref{eq:chi})),
  where $n$ is the upstream density, and $\urad$ is the energy density of the soft (optical) radiation.  Left and right panels correspond to assumed values of the postshock magnetization of $\epsB=10^{-6}$ and $\epsB=10^{-4}$, respectively.  The fluxes are computed assuming fast cooling for both thermal and non-thermal processes and we have adopted characteristic values for the non-thermal injection fraction $\epsnth=10^{-2}$, injection index $q=2$, maximal energy of accelerated electrons $\gamma_{\rm max} = 10^5$,  and optical luminosity $\Lrad = 10^{38}$~erg~s$^{-1}$ ($\urad = \Lrad/(4\pi c R^2)$), where $R = \vsh t$ and time $t = 1$~week.
  However, note that the isocontours of $\nu F_{\nu, 30 \mathrm{keV}}/\nu F_{\nu, 1 \mathrm{GeV}}$ are independent of $\Lrad$ and $t$, and depend weakly on $\epsnth$.
  Black dashed lines show isocontours of constant upstream density for the chosen $\Lrad$ and $t$,  given by
  $\chi\propto \Lrad/(\vsh^2 t^2)$.  Red dashed lines show isocontours of constant gamma-ray to optical flux ratios, $\nu L_{\nu, 1 \mathrm{GeV}}/\Lrad$, which scale linearly with $\epsnth$ but are independent of $\Lrad$ and $t$.  The region to the right of the black dotted lines ($\Lsh > \Lrad$) is unphysical as the total shock-generated luminosity (a large fraction of which is absorbed and reprocessed to optical frequencies) cannot exceed $\Lrad$.
}
  \label{Fig:ratio:lept}
\end{figure*}

\begin{figure} 
  \begin{center}
  \begin{tabular}{c}
  \includegraphics[width=0.45\textwidth]{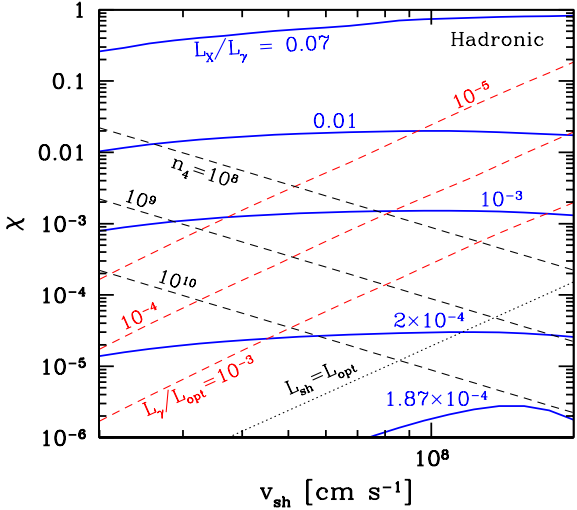}
  \vspace*{-0.4cm}
  \end{tabular}
  \end{center}
  \caption{Same as Figure \ref{Fig:ratio:lept}, but for hadronic models.
  Parameters: 
  fraction of shock energy injected into nonthermal protons $\epsp=0.1$, injection slope $\qp=2$, where 
  $d\Np/d(\gp\bp) \propto (\gp\bp)^{-\qp}$,
  optical luminosity $\Lrad= 10^{38}$~erg~s$^{-1}$,
  postshock magnetization $\epsB=10^{-6}$.
  The scalings of the different isocontours are the same as for the leptonic models, except $\epsp$ replaces $\epsnth$.
  }
  \label{Fig:ratio:hadr}
\end{figure}

The dimensionless ratio, $\chi$, of the radiation compactness and Thomson optical depth (Equation~\ref{eq:chi}) controls the partitioning of the non-thermal energy emitted
in the hard X- and gamma-ray bands.  Both thermal and non-thermal particles cool at a rate which is proportional to
either the radiation energy density $\urad$ (IC) or matter density $n$ (line cooling, bremsstrahlung, Coulomb, pp-collisions);
thus $\chi\propto \urad/n$ determines their relative importance.

Figures \ref{Fig:ratio:lept} and \ref{Fig:ratio:hadr} show contours of constant $L_{\rm X}/L_{\gamma}$
in the $\vsh-\chi$ plane for leptonic and hadronic scenarios, respectively.
The most obvious trend is that higher values of $L_{\rm X}/L_{\gamma}$ are obtained at higher $\chi$.
This can be understood from Equations (\ref{eq:Ebr}) and (\ref{eq:EIC}), which show that IC becomes more dominant as $\chi$ increases.
Once IC dominates,
the flux in the hard X-ray band is roughly proportional to $\chi$ for a fixed total non-thermal energy (i.e., constant $Q(\gamma)$ in Equation (\ref{eq:EIC})).
In simple terms, for a given electron energy $\gamma$,
most of the IC power is emitted at lower frequencies ($x\approx \gamma^2 x_0$) compared to bremsstrahlung ($x\sim \gamma$),
thus resulting in stronger X-ray emission
in the former case.
In the limit of complete IC dominance, the leptonic model spectrum approaches the fast-cooling shape of $\nu F_{\nu} \propto \nu^{(2-q)/2}$, i.e. flat for $q=2$.
However, note that Equation (\ref{eq:EIC}) is no longer valid in this limit since Coulomb losses also become negligible at high $\chi$.

Another key feature is the existence of a lower limit of $L_{\rm X}/L_{\gamma} \gtrsim 10^{-3}$ in the leptonic case and $\gtrsim 10^{-4}$ in the hadronic case.
This corresponds to the complete dominance of bremsstrahlung over IC losses attained at low $\chi$.
Due to the low-energy tail of the bremsstrahlung spectrum following $\nu  F_{\nu} \propto \nu$,
the relative power emitted in the X-ray and gamma-ray bands by leptonic emission must exceed $\nu_{\rm X}/\nu_{\gamma} \sim 10^{-3}$.
In the leptonic scenario, the lower limit on $L_{\rm X}/L_{\gamma}$ is actually somewhat higher than this
because the X-ray emission receives additional contributions from electrons with energies $\gamma\lesssim 10^2$ which are too low to contribute in the gamma-ray band.

In the hadronic scenario, most of the gamma-ray flux arises from the decay of neutral pions; however,
an appreciable contribution also comes from the IC and bremsstrahlung emission from $\epm$ pairs injected by $\pi_{\pm}$ decay.
However, in contrast to the leptonic case, the decay of charged pions creates very few pairs below $\gamma\sim 10^2$, which would contribute to the X-ray flux.
As a result, the minimum value of $L_{\rm X}/L_{\gamma} \sim 10^{-4}$ is roughly an order of magnitude lower than in the leptonic case.

An independent constraint on the parameter space is obtained from the gamma-ray to optical flux ratio, as shown by red dashed lines in Figures \ref{Fig:ratio:lept} and \ref{Fig:ratio:hadr}.
For a given observed value of $L_{\gamma}/\Lrad$, the allowed region on the $\vsh - \chi$ plane lies between the black dotted line and the corresponding red dashed line.
Combined with a measurement of $L_{\rm X}/L_{\gamma}$, this could in principle be used to lift
the degeneracy between $\vsh$ and $\chi$ (or, equivalently, the density $n$) and to make an estimate of both.
However, this assumes that the other uncertain parameters $\epsnth$ ($\epsp$), $\epsB$ are known or can be constrained.  

In fact, the existence of an allowed parameter region for a measured $L_{\gamma}/\Lrad$ sets
a lower limit on the particle acceleration efficiency \citep{Metzger+15}.  For the flat acceleration spectra assumed in Figures \ref{Fig:ratio:lept} and \ref{Fig:ratio:hadr}  ($q,\qp=2$),
the efficiencies $\epsnth$ or $\epsp$ are constrained to similar values in the two scenarios
(note that we assume $\epsnth=10^{-2}$ in Figure \ref{Fig:ratio:lept}, while $\epsp=0.1$ in Figure \ref{Fig:ratio:hadr}; $\Lg/\Lrad$ scales approximately $\propto \epsnth$, $\epsp$).
As pointed out by \citet{Metzger+15}, high observed values of $\Lg/\Lrad$ (e.g. $\sim 10^{-2}$ in Nova V1324 Sco and $\sim 3\times 10^{-4}$ in V399 Del)
favor hadronic scenarios on both theoretical and observational grounds.  Both particle-in-cell plasma simulations (e.g.~\citealt{Kato15}, \citealt{Park+15})
as well as modeling of observed supernova remnants (e.g.~\citealt{Morlino&Caprioli12}), suggest a relatively low electron acceleration efficiency of $\epsnth \lesssim 10^{-3}$ in non-relativistic shocks.

\section{Constraints on mass loss rate and density}

\label{sec:Mdotn}

\begin{figure*}[h]
  \begin{center}
  \begin{tabular}{cc}	
  \includegraphics[trim = 0cm 0cm 0cm 0cm, width=0.45\textwidth]{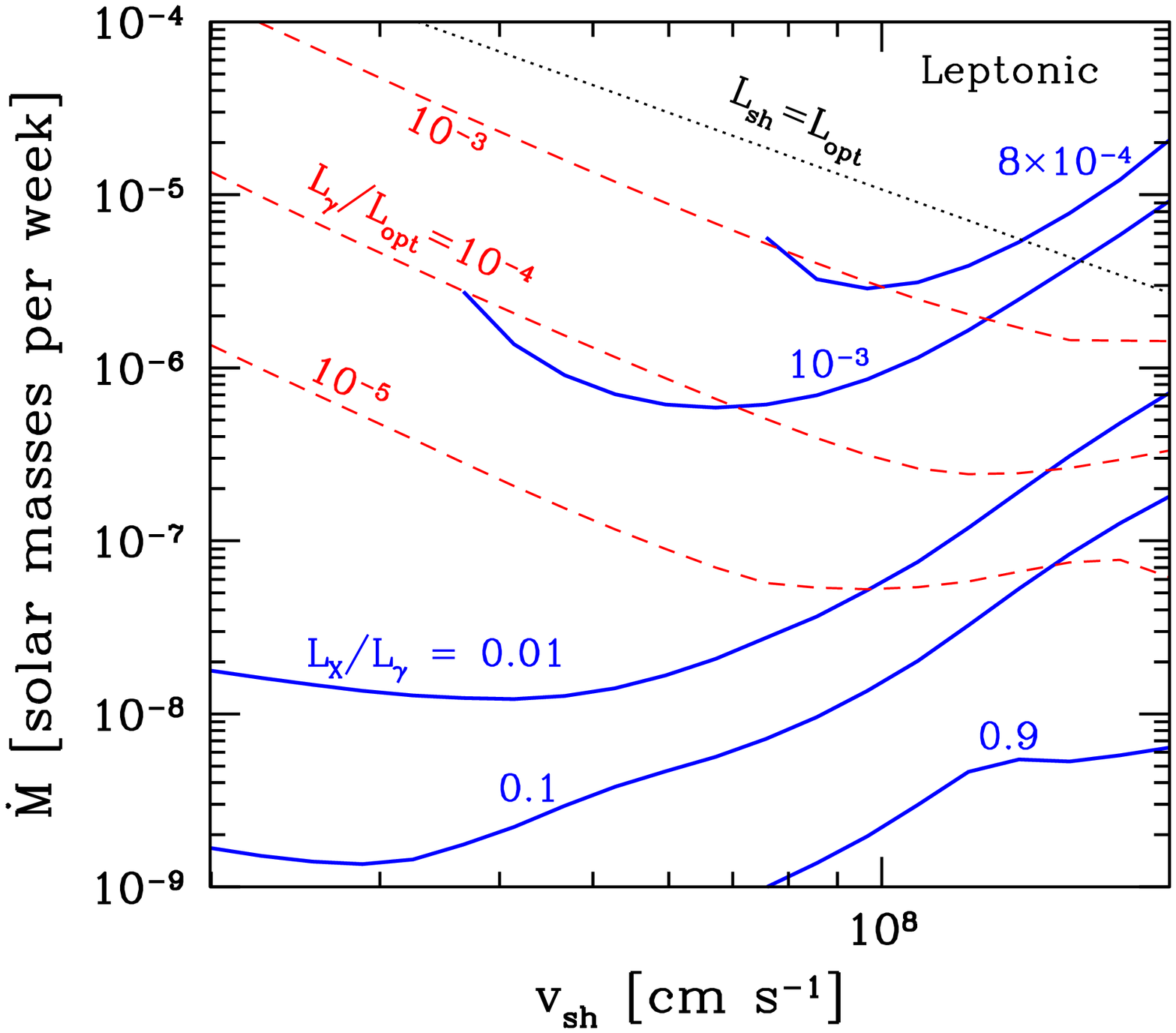}	&
  \includegraphics[trim = 0cm 0cm 0cm 0cm, width=0.45\textwidth]{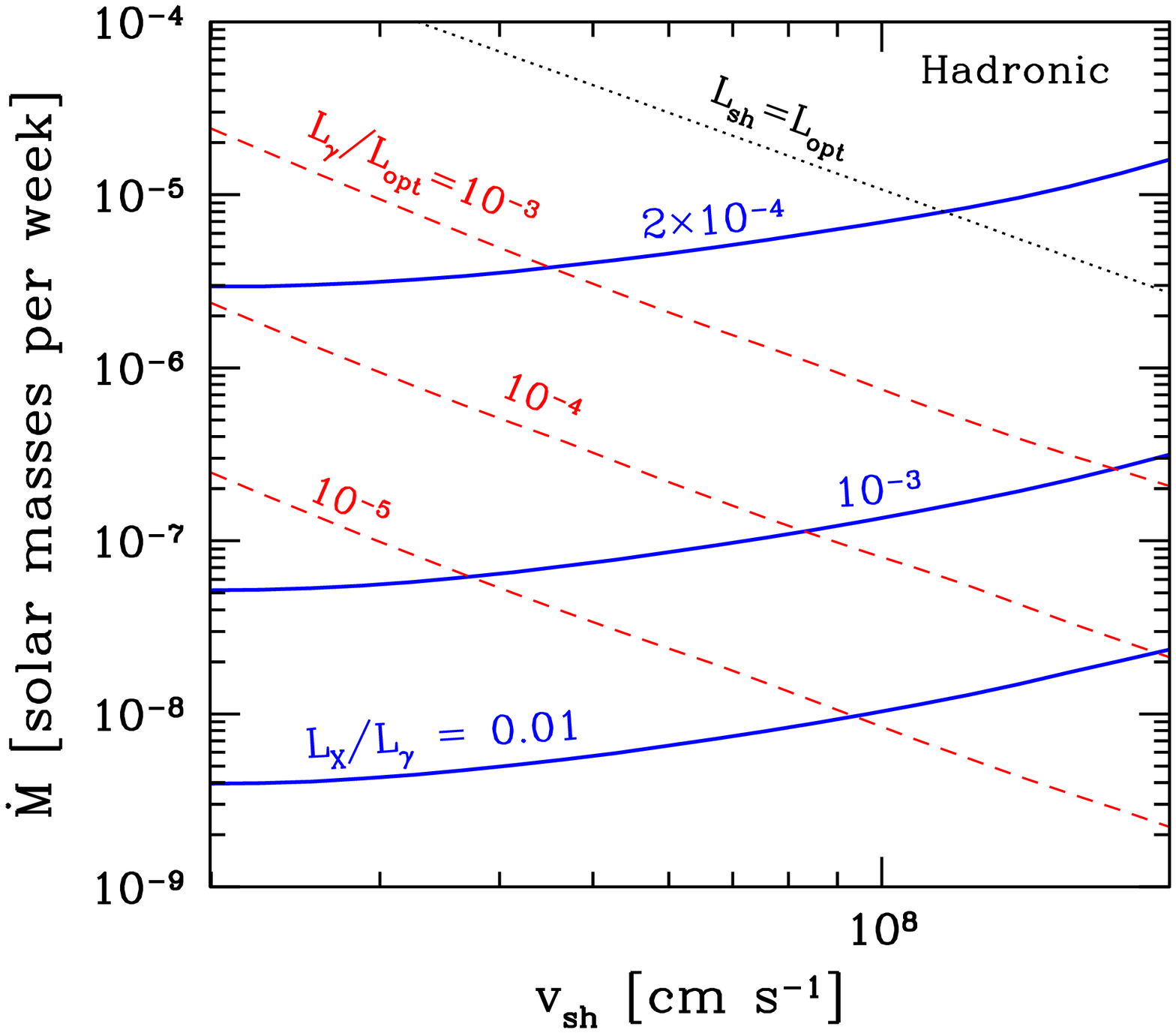}
  \end{tabular}
  \end{center}
  \vspace*{-0.4cm}
  \caption{
  Isocontours of 
  relative fluxes at $30$~keV and $1$~GeV (blue solid lines) in $\vsh-\Mdot$ space, for leptonic (left panel) and hadronic models (right panel).
  Isocontours of constant
  $\nu L_{\nu, 1 \mathrm{GeV}}/\Lrad$
  are shown by red dashed lines.
  The allowed parameter region lies to the left of the black dotted line, where $\Lsh < \Lrad$.
  Parameters: postshock magnetization $\epsB=10^{-6}$, optical luminosity $\Lrad=10^{38}$~erg~s$^{-1}$, $\zeta = \vsh/\vw = 1$.
  Leptonic model (left): non-thermal injection fraction $\epsnth=10^{-2}$, injection index $q=2$, maximal Lorentz factor of accelerated electrons $\gamma_{\rm max} = 10^5$.
  Hadronic model (right): injection fraction $\epsp=0.1$, index $\qp=2$, maximal proton Lorentz factor $\gpmax = 10^3$. 
  The contours shift vertically proportionally to $\Mdot \propto \Lrad/\zeta$ if $\Lrad$ or $\zeta$ is varied.
  }
  \label{fig:ratio:Mdot}
\end{figure*}

The allowed region in
$\vsh-\chi$ parameter space obtained from simultaneous gamma-ray and hard X-ray observations
can be used to constrain
the mass outflow rate of the ejecta, as well as the density at the shock if one has an independent handle on the shock radius.

In novae,
one can visualize two scenarios for the shock formation,
(1) a fast wind from the central object impacting upon
dense ``external'' shell (possibly ejected at an earlier phase of the same nova eruption),
and/or
(2) internal shocks within a single variable outflow.
In both cases, one can express $\chi$ as
\begin{align}
\chi = 
\frac{\mprot}{\me} \frac{\Lrad \vw}{\Mdot c^3} = 
2.1 \times 10^{-5} \, \frac{L_{{\rm opt}, 38} \, v_{{\rm w},8}}{ \dot{M}_{-5}},
\label{eq:chi:Mdot}
\end{align}
where $\vw$ is the outflow velocity and $\dot{M}_{-5}$ is the mass outflow rate in units of $10^{-5}M_{\odot}$ per week (a typical value in novae).
It is important to note that Equation (\ref{eq:chi:Mdot}) does not explicitly depend on $R$, as the main unknown parameters that control $L_{\rm X}/L_{\gamma}$ are
$\Mdot$ and $\vw$ in the shock upstream, and the shock velocity $\vsh$.

In either scenario,
$\vw$ is unlikely to be very different from the shock velocity $\vsh$.
In case of internal shocks,
dissipating a substantial fraction of the outflow energy (as suggested by gamma-ray observations) requires $\vsh \sim \vw$.
In the case of a fast tenuous wind impacting a slow dense shell most of the energy is dissipated at the reverse shock,
i.e. the shock running back into the wind material;
as long as the velocity contrast between the two media is substantial, on again finds $\vsh \sim \vw$
We will therefore parametrize
$\vsh = \zeta \vw$, with $\zeta \lesssim 1$.

The isocontours of $L_{\rm X}/L_{\gamma}$ in $\vsh-\Mdot$ space are shown in Figure \ref{fig:ratio:Mdot}.
As expected, the X-ray to gamma-ray luminosity ratio decreases with increasing mass outflow rate, which results in higher densities and stronger bremsstrahlung and Coulomb losses relative to IC.

To date,
there has been no unambiguous detection of non-thermal X-rays from novae.
Simultaneous hard X-ray and gamma-ray observations have been performed in two events,
V339 Del and V5668 Sgr (Mukai et al., in prep).
Both novae were detected by {\it Fermi}/LAT days to weeks after the optical outburst (\citealt{Ackermann+14}, \citealt{Cheung+16}).
The {\it NuSTAR} satellite
observed V339 Del (24 ks) and V5668 Sgr (52 ks) approximately 1 and 2 weeks after the onset, respectively.
The upper limits for the $20$~keV flux were obtained  (Mukai et al.~, in preparation):
$\nu F_{\nu} < 1.2 \times 10^{-13}$~erg~cm$^{-2}$~s$^{-1}$ (V339 Del) and
$\nu F_{\nu} < 3.5 \times 10^{-14}$~erg~cm$^{-2}$~s$^{-1}$ (V5668 Sgr).
Comparison with the simultaneous LAT fluxes yielded the ratio of $20$~keV to $100$~MeV fluxes/luminosities:
$L_{\rm X}/L_{\gamma} < 4.0\times 10^{-3}$ (V339 Del) and	
$L_{\rm X}/L_{\gamma} < 1.7\times 10^{-3}$ (V5668 Sgr). 	

The optical fluxes at the time of the X- and gamma-ray observations were approximately
$10^{-7}$~erg~cm$^{-2}$~s$^{-1}$ (V339 Del) and
$6\times 10^{-7}$~erg~cm$^{-2}$~s$^{-1}$ (V5668 Sgr);
the corresponding gamma-ray to optical flux ratios were $\sim 3\times 10^{-4}$ and $\sim 3\times 10^{-5}$,
respectively (\citealt{Skopal+14}, \citealt{Metzger+15}, \citealt{Munari+15}).

\begin{figure*} 
  \begin{center}
  \begin{tabular}{cc}	 
  \includegraphics[trim = 0cm 0cm 0cm 0cm, width=0.45\textwidth]{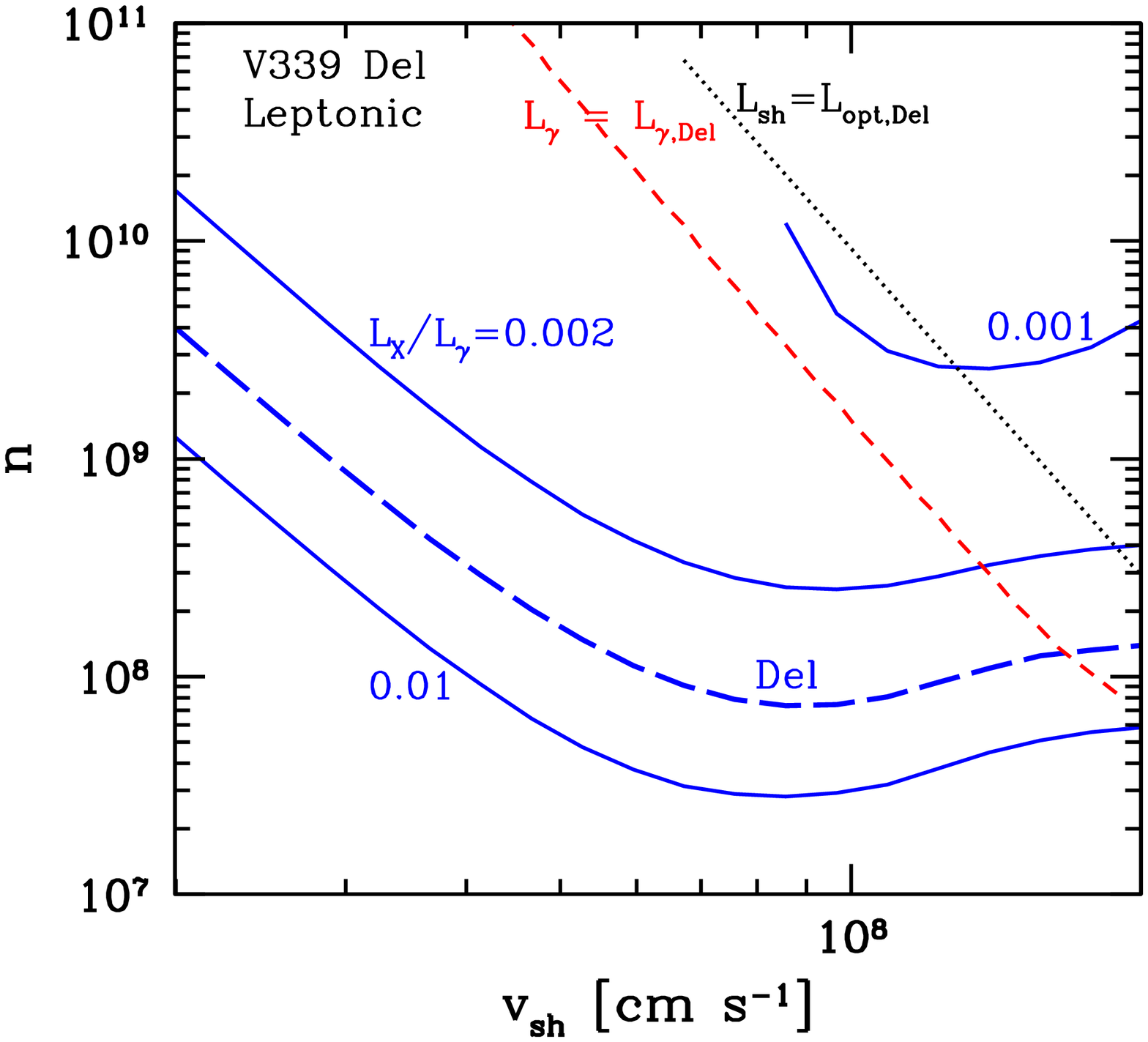}	&
  \includegraphics[trim = 0cm 0cm 0cm 0cm, width=0.45\textwidth]{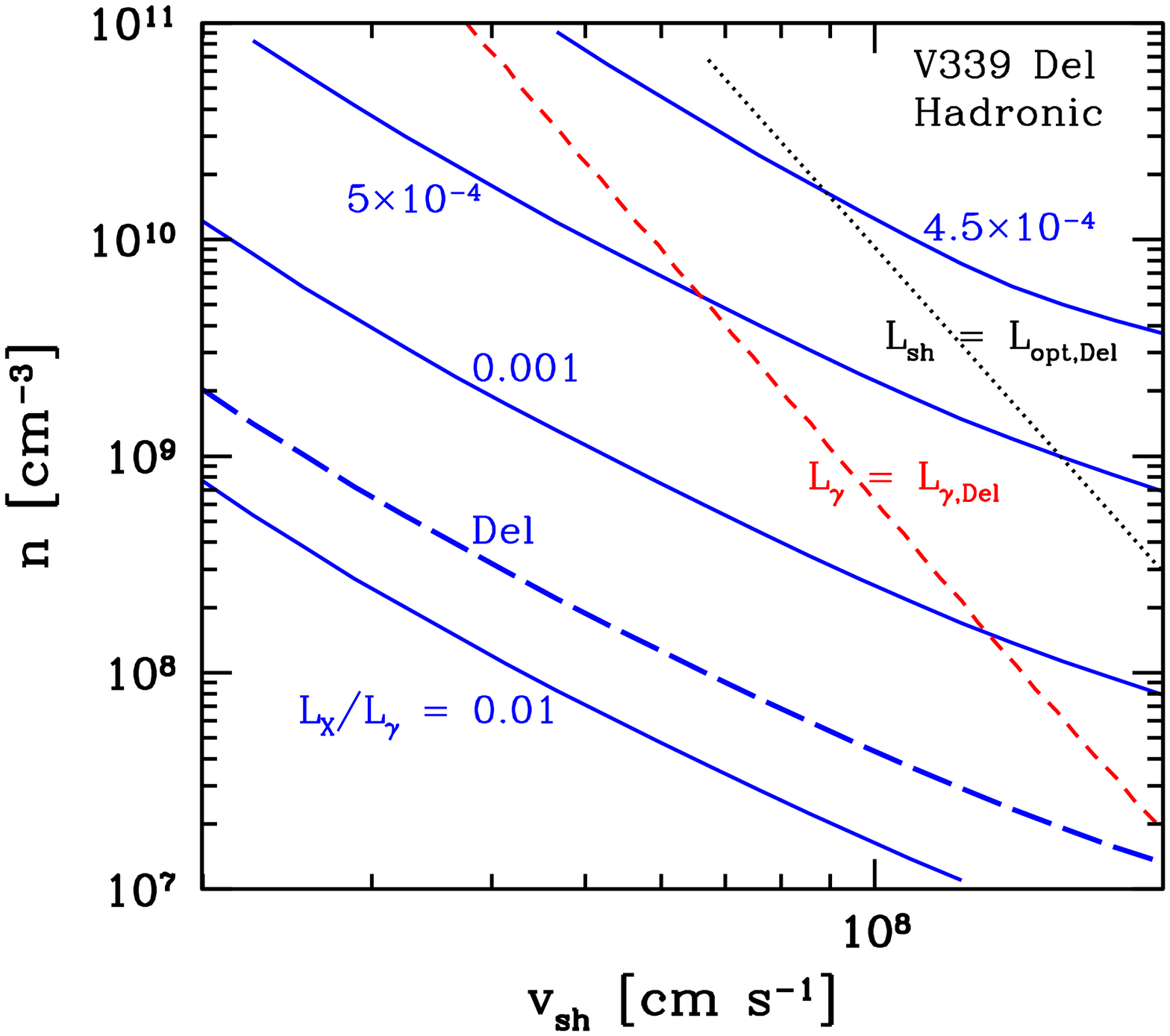}
  \end{tabular}
  \end{center}
  \vspace*{-0.4cm}
  \caption{
  Isocontours of relative luminosities at $20$~keV and $100$~MeV for nova V339 Del (blue solid lines), for leptonic (left panel) and hadronic models (right panel);
  the bold long-dashed line corresponds to the observed upper limit.
  Red dashed line: isoline corresponding to the observed gamma-ray luminosity
  $\nu L_{\nu, 100 \mathrm{MeV}} \approx 6\times 10^{34}$~erg~s$^{-1}$ (\citealt{Ackermann+14}),
  assuming a distance $d=4.2$~kpc.
  Black dotted line: $\Lsh = \Lrad$.
  Parameters: $\Lrad=2\times 10^{38}$~erg~s$^{-1}$, shock radius $R=\vsh t$, where $t=1$~week;
  other parameters the same as in Figure \ref{fig:ratio:Mdot}:
  $\epsB=10^{-6}$, $\epsnth=10^{-2}$, $q=2$, $\gamma_{\rm max} = 10^5$,
  $\epsp=0.1$, $\qp=2$, $\gpmax = 10^3$.
  }
  \label{Fig:ratio:Del}
\end{figure*}

\begin{figure*} 
  \begin{center}
  \begin{tabular}{cc}
  \includegraphics[trim = 0cm 0cm 0cm 0cm, width=0.45\textwidth]{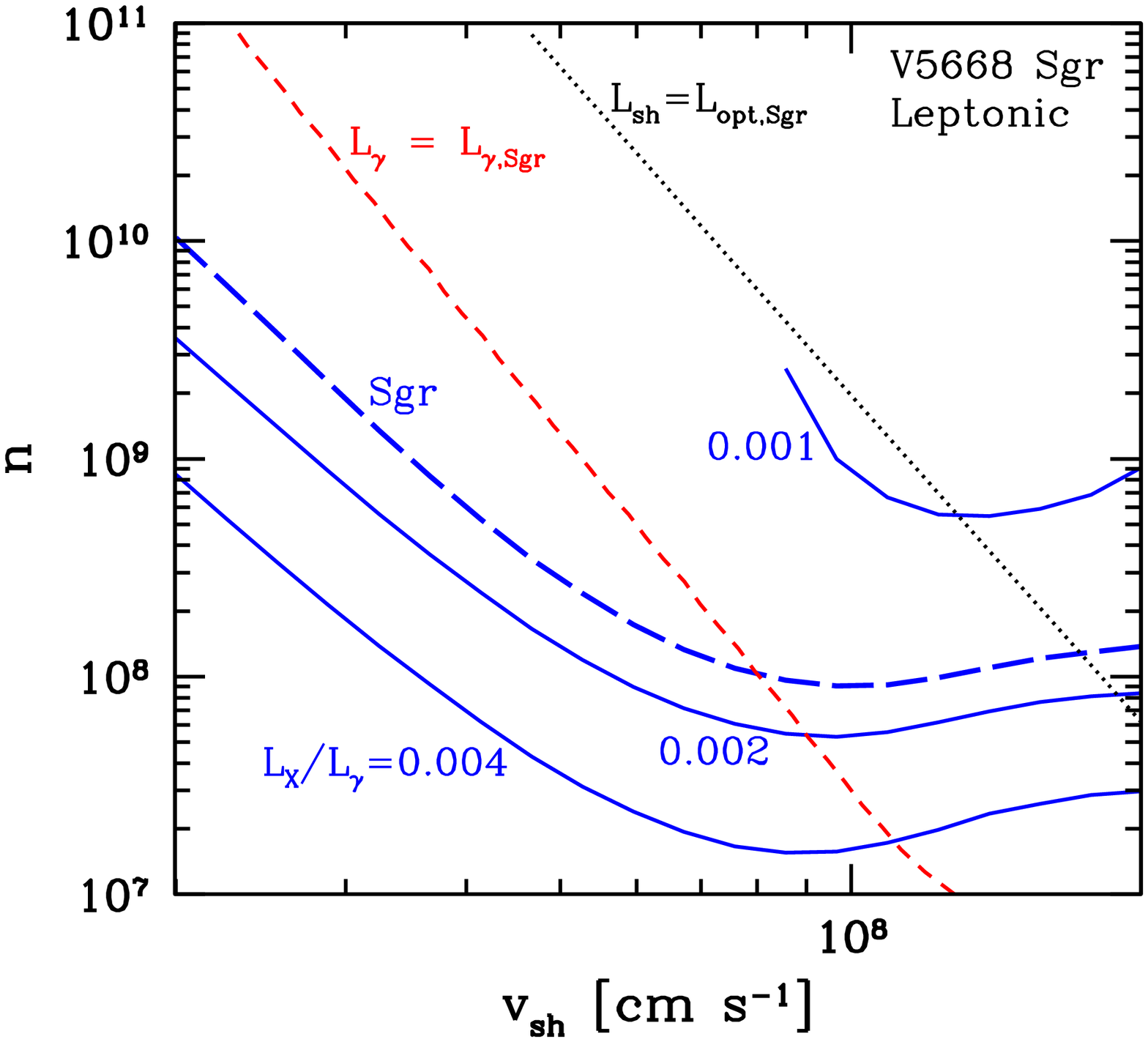}	&
  \includegraphics[trim = 0cm 0cm 0cm 0cm, width=0.45\textwidth]{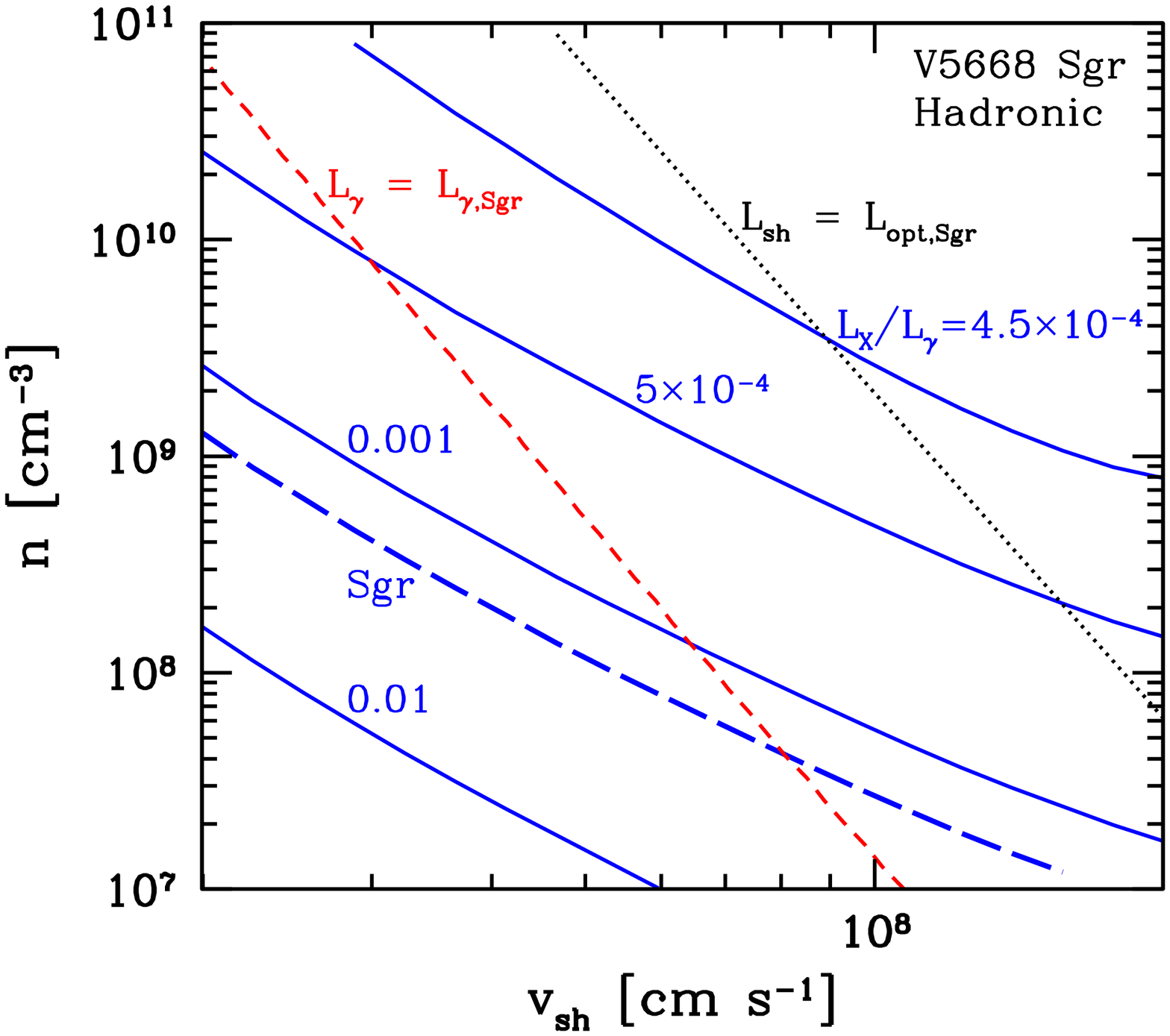}	
  \end{tabular}
  \end{center}
  \vspace*{-0.4cm}
  \caption{
  Isocontours of relative luminosities at $20$~keV and $100$~MeV for nova V5668 Sgr (blue solid lines), for leptonic (left panel) and hadronic models (right panel);
  the bold long-dashed line corresponds to the observed upper limit.
  Red dashed line: isoline corresponding to the observed gamma-ray luminosity
  $\nu L_{\nu, 100 \mathrm{MeV}} \approx 6\times 10^{33}$~erg~s$^{-1}$ (\citealt{Cheung+16}),
  assuming a distance $d=1.5$~kpc (\citealt{Banerjee+16}).
  Black dotted line: $\Lsh = \Lrad$.
  Parameters: shock radius $R=\vsh t$, where $t=2$~weeks, $\Lrad=1.7\times 10^{38}$~erg~s$^{-1}$;
  other parameters the same as in Figure \ref{Fig:ratio:Del}.
  }
  \label{Fig:ratio:Sgr}
\end{figure*}

The theoretical $L_{\rm X}/L_{\gamma}$ isocontours on the $\vsh-n$ plane
for nova V339 Del and V5668 Sgr are shown in Figures \ref{Fig:ratio:Del} and \ref{Fig:ratio:Sgr}, respectively.
The allowed region as determined by the gamma-ray and optical observations
lies between the red dashed and black dotted lines.
Unfortunately, the present X-ray upper limits do not yield significant additional constraints on the parameter space in either nova.
The leptonic case is somewhat more constraining,
requiring $\vsh \lesssim 2\times 10^{8}$~cm~s~$^{-1}$ and $n \gtrsim 10^{8}$~cm$^{-3}$ in both novae,
which incidentally ensures that the shocks are radiative (see Figure \ref{Fig:cool}).

It is worth noting that the {\it NuSTAR} limit for V5668 Sgr is sufficiently deep to give hope for more interesting constraints in future events.  The LAT fluxes in several gamma-ray novae have exceeded
 $10^{-10}$~erg~cm$^{-2}$~s$^{-1}$ (\citealt{Ackermann+14}),
 which could yield X- to gamma-ray ratios of the order $\sim 3\times 10^{-4}$.
 This is sufficient to either result in a detection or
 rule out leptonic models.
Unfortunately, nova V5668 Sgr was intrinsically about an order of magnitude weaker in gamma rays compared to more luminous events such as V1324 Sco (\citealt{Cheung+16}, their Fig.~5).

\subsection{Escape of the shock radiation from the outflow}

\begin{figure} 
  \begin{center}
  \includegraphics[trim = 0cm 0cm 0cm 1cm, width=0.45\textwidth]{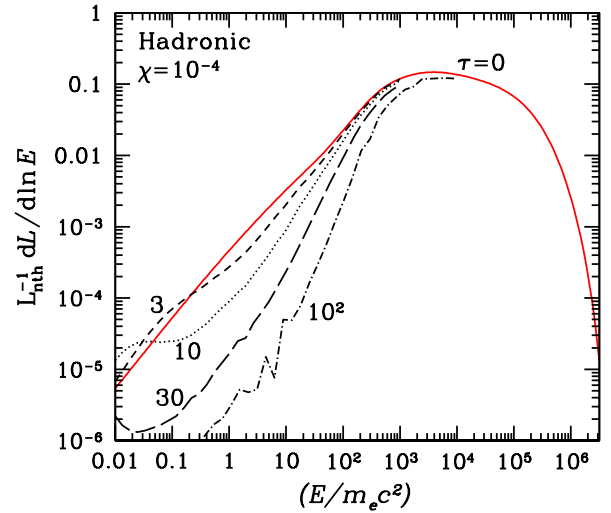}
  \end{center}
  \vspace*{-0.4cm}
  \caption{Shock spectrum from Figure \ref{Fig:sp:hadr} (red solid line),
  after diffusing out of a wind-like outflow (black lines).
  Different lines correspond to different Thomson optical depths of the shock within the wind.
  }
  \label{fig:screen}
\end{figure}

Our discussion so far has
considered only the {\it intrinsic} emission
from radiative shocks and assumed that the generated X-rays and gamma-rays can freely escape from the ejecta.
However, if the shocks
take place sufficiently deep in the outflow,
the dense ambient material can
leave a strong imprint on the escaping radiation.
Above a few tens of keV,
the main source of opacity in the outflow is Compton scattering.
At $E\ll \me c^2$,
the average fractional energy loss of a photon in a scattering event is $\sim x = E/\me c^2$.
Emitted at $\tauT\gg 1$, the photon experiences approximately $\tauT^2$ scatterings before escaping.
Thus if $x \tauT^2 \gtrsim 1$, 
or equivalently $\tauT\gtrsim 5 \,(E/20\mbox{keV})^{-1/2}$,
the photon energy is significantly degraded as it diffuses out of the ejecta.
This can be seen in Figure \ref{fig:screen}:
as $\tauT$ is increased,
the soft gamma-rays ($E\sim 1$~MeV) are depleted first, followed by hard X-rays at progressively lower energies.
Note that if the primary spectrum is sufficiently hard
($F_{\nu}\propto \nu^{-\alpha}$ with $\alpha<0$),
the emission in a given band is initially enhanced as $\tauT$ is increased,
at the expense of higher-energy photons being downscattered into the band,
before being suppressed at higher $\tauT$.

At $E \gg \me c^2$, the Klein-Nishina cross-section approximately follows $\sigma_{\rm KN}\approx(3\sigmaT/8x)\ln(x)$,
while a photon loses most of its energy in a single scattering event.
At 100~MeV, $\tau_{\rm KN} \approx \tauT/100$,
i.e. the photons in the LAT band suffer significant recoil losses
if $\tauT\gtrsim 100$ (Figure \ref{fig:screen}).

Note that high $\tauT$ also has the effect of enhancing the (optical) radiation density in the ejecta,
as $\urad\approx \Lrad(1+\tauT)/(4\pi c R^2)$.
The $\chi$ parameter is enhanced by the same factor $(1+\tauT)$,
which has a positive effect
on the hard X-ray emission (Figures \ref{Fig:ratio:lept} and \ref{Fig:ratio:hadr}).

In summary,
for given shock parameters,
the X-ray to gamma-ray ratio is somewhat increased if $\tauT \sim$~a~few.
In the range $\tauT \approx 10-100$ the X-rays are strongly suppressed while LAT gamma rays still escape unhindered.
At even higher opacities
the GeV gamma-rays are also significantly degraded.

\section{Discussion and conclusions}

\label{sec:concl}

Non-thermal emission from non-relativistic shocks provides a wealth of 
information about the shock environment as well as the physics of particle acceleration.
In dense media characteristic of e.g. nova eruptions during the first weeks, 
the heated plasma rapidly cools and compresses behind the shock due to line-cooling and bremsstrahlung emission.
The unique property of slow ($\vsh\lesssim 10^8$~cm~s$^{-1}$) shocks is that
the cooling time of the relativistic particles responsible for the hard X-ray and gamma-ray emission
is longer than the thermal cooling/compression time.
Thus the high-energy radiation samples a range of physical conditions behind the shock front,
where the density as well as magnetization can change by a few orders of magnitude.

In this work we have computed the non-thermal emission from the cooling layer behind the shock front,
due to relativistic bremsstrahlung and IC upscattering of thermal (optical) radiation,
as well as hadronic collisions leading to both gamma-ray and $\epm$ production.
Additional losses due to Coulomb collisions with the thermal plasma and synchrotron cooling were also taken into account.
The downstream compression was calculated using a simple prescription
by keeping the total pressure constant,
and explicitly following the evolution of partial pressures of the thermal and non-thermal plasma as well as the magnetic field
as the plasma cools.

The focus of our analysis was on using spectral information from flux ratios in different bands to constrain
the physical conditions at the shock as well as particle acceleration mechanisms.
In particular, we concentrated on the hard X-ray and GeV gamma-ray bands accessible to {\it NuSTAR} and {\it Fermi}/LAT, respectively.
In contrast to soft X-rays,
hard X-ray radiation is more likely to be representative of the intrinsic non-thermal emission from the shock:
it is not impeded by bound-free absorption and can only be degraded by Compton recoil if $\tauT \gtrsim 10$.
Furthermore, the hard X-ray band is less likely to be contaminated by free-free emission
from shock-heated thermal electrons, except if $\vsh\gtrsim 2\times 10^8$~cm~s$^{-1}$.

In radiative shocks, the gamma-ray output above a few hundred MeV
roughly traces the total energy placed into $>1$~GeV particles, even though the detailed spectrum depends on the particular model of particle acceleration as well as the dominant radiative process.
In contrast, emission at lower frequencies is more sensitive to the physical conditions near the shock.
If the shock is embedded in an external radiation field with luminosity exceeding the shock-dissipated power
(usually expected in novae),
the fraction of non-thermal energy emerging in the X-ray band
anticorrelates with density.

At very low densities ($\chi = \lrad/\tauT \sim 1$, see Equation \ref{eq:chi})
IC cooling dominates the electron (positron) energy loss, and comparable energy is radiated in the X-ray and gamma-ray bands by leptonic emission.
This regime corresponds to low shock power.
At high densities ($\chi \lesssim 10^{-4}$),
relativistic bremsstrahlung and Coulomb collisions are the dominant cooling mechanisms for $\lesssim 1$~GeV electrons.
In this regime,
the hard X-ray emission is a superposition of the low-energy tails of the bremsstrahlung spectra
from relativistic leptons between $\gamma \approx$ a few to $10^3$, attenuated by Coulomb losses.
The high density/low $\chi$ regime corresponds to high shock power
and is therefore most relevant for practical (detection) purposes,
however the ratio of X-ray to gamma-ray energies is relatively low in this case, $L_{\rm X}/L_{\gamma} \approx 10^{-4} - 10^{-3}$.
In the extreme high density limit the X-ray to gamma-ray ratio approaches an asymptotic value in both leptonic and hadronic scenarios,
which is approximately three times higher in the leptonic case.

\subsection{Gamma-ray novae}

There is mounting evidence that strong shocks are
commonplace in classical novae,
which provide an independent
avenue of constraining the the
properties of nova outflows.
Simultaneous {\it Fermi}/LAT and optical observations
of e.g. V1324 and V399 Del
already strongly limit the allowable parameter space (\citealt{Metzger+15});
in particular, they place
a lower limit on the shock luminosity,
which can be written
as $\Lsh = (9/32) \zeta \Mdot \vsh^2$, where $\zeta=\vsh/\vw$.
We have shown that
the degeneracy between $\Mdot$ and $\vsh$ can be lifted,
at least in principle,
by a concurrent hard X-ray observation.
The relevant observational measure is the ratio of X-ray and gamma-ray fluxes,
which places an independent constraint on the allowed region $\vsh-\Mdot$ space,
without explicit reference to e.g. the shock radius or geometry.

Unfortunately, the presently available {\it NuSTAR} upper limits for two classical novae, V339 Del and V5668 Sgr,
are not sufficiently deep
to yield significant constraints, given their gamma-ray fluxes.
There is reason for optimism, however,
since the flux limits attainable by a $\sim 50$ ks {\it NuSTAR} observation
(as performed for V5668 Sgr)
of novae with higher gamma-ray fluxes such as e.g. V1324 Sco or V959 Mon
would start pushing the theoretical limit of the $L_{\rm X}/L_{\gamma}$ ratio,
and likely result in a detection.
Failing that, a deep upper limit could
still be useful by ruling out leptonic models,
for which $L_{\rm X}/L_{\gamma} \gtrsim 5\times 10^{-4}$ for any reasonable parameters.  

A low X-ray luminosity could instead result from attenuation due to inelastic electron scattering by a high column of gas ahead of the shock with optical depth $\tauT \gtrsim 5$, in which case constraints on the shock properties from an X-ray non-detection would not be as strong; however, for $\tauT \gtrsim 100$ the 100 MeV gamma-ray emission would itself be blocked, thus limiting the range of $\tauT$ over which this explanation would be viable to roughly one order of magnitude.  

The model presented in this paper assumes a 1D planar shock, and constant post-shock pressure.
These simplifications may be questionable given the highly multi-dimensional thermal and thin-shell instabilities known to plague radiative shocks
(e.g.~\citealt{Vishniac83}, \citealt{Chevalier&Imamura82}). 
Nevertheless, insofar as the local thermodynamic conditions experienced by a cooling parcel of thermal and relativistic particles
are reasonably captured by the simple processes of cooling and compression described here, these complications should not impact the qualitative features of our results.

\subsection{Colliding wind binaries}

The colliding stellar winds of early-type stars 
(O, B, Wolf-Rayet) in binary systems
give rise to strong shocks that
can accelerate both electrons and protons to high energies (\citealt{DeBecker07}, and references therein).
The existence of relativistic particles in these systems has been proven by the detection of radio synchrotron emission (e.g., \citealt{Abbott+86}; \citealt{Chapman+99}).
To date, no gamma rays have been detected in CWB
(with the possible exception of $\eta$ Carinae; \citealt{Hamaguchi+14}; \citealt{Reitberger+15});
upper limits for a sample of 7 systems have been obtained by {\it Fermi}/LAT (\citealt{Werner+13}).
Non-thermal X-rays have not been detected yet with {\it Integral} (e.g.~\citealt{deBecker+07}),
although \citet{Sugawara+11} present {\it Suzaku} observations showing evidence for a hard power-law X-ray component in WR140.

In tight binaries such as WR20a, or near the periastron passage of eccentric systems (e.g. WR 140)
the particle densities at the shock
are comparable to those expected in novae,
and the shocks may become radiative.\footnote{Evidence for radiative shocks is provided by observed dust formation between
the colliding wind shocks of WR binaries (e.g.~\citealt{Williams+12}), thus indicating the presence of cold and neutral gas.}
The shock/wind velocities are also similar, 
typically $0.3-6\times 10^8$~cm~s$^{-1}$ (see e.g. \citet{Crowther07} for a review).
However, the mass outflow rate, $\Mdot \sim 10^{-4}-10^{-5} \, \mbox{M}_{\sun}$~yr$^{-1}$ is typically somewhat lower than in classical nova eruptions.
Combined with comparable or higher optical/UV luminosities, $L_{\rm opt} \sim 10^{5} - 10^{6} \, \mbox{L}_{\sun}$,
the cooling regime of the relativistic particles differs from novae.
This can be seen by writing the $\chi$ parameter as
\begin{align}
\chi = 0.04  \, v_8 \, \left(\frac{L_{\rm opt}}{10^6\,\mbox{L}_{\sun}}\right) \left( \frac{\Mdot}{10^{-5} \, \mbox{M}_{\sun} \mbox{yr}^{-1}} \right)^{-1}.
\end{align}
Recalling Figure \ref{Fig:cool} (left panel), one concludes that the relativistic leptons cool predominantly by IC emission rather than bremsstrahlung (or Coulomb);
furthermore, the IC cooling of the hard X-ray and gamma-ray emitting electrons is typically
{\it faster} than the cooling of the thermal plasma (Equation \ref{eq:tICrel}).
Therefore, in leptonic models the X-ray and gamma-ray emissivities are not sigificantly affected
by downstream compression, nor are the X-rays necessarily suppressed by Coulomb losses.
As a result, the energy emitted in the hard X-ray and gamma-ray bands can be comparable (Figure \ref{Fig:ratio:lept}).

In the hadronic scenario, the cooling time via pp-collisions relative to the compression time
depends only on the shock velocity (Equation (\ref{eq:tpprel}) and Figure \ref{Fig:cool}, right panel).
If $\vsh \lesssim 3\times 10^{8}$~cm~s$^{-1}$, the accelerated protons deposit most of their energy only
after the thermal pressure has been lost and the downstream plasma has significantly compressed.
This mainly affects the X-ray emission from secondary $\epm$ pairs from $\pi_{\pm}$ decay,
which can experience both synchrotron losses in the compression-enhanced magnetic field,
as well as increased bremsstrahlung losses that give rise to harder spectra.
Therefore the X-ray to gamma-ray ratio is expected to be
over an order of magnitude lower than in the leptonic case.

The {\it Fermi} upper limits (\citealt{Werner+13}) are at odds with
theoretical predictions for the gamma-ray flux by several groups (\citealt{Reimer+06}; \citealt{Benaglia&Romero03}; \citealt{Pittard&Dougherty06}).
The discrepancy has not yet been resolved.
Strong synchrotron losses could provide a possible explanation
if the downstream plasma is able to compress and amplify the magnetic field
before the relativistic particles have cooled.
This however requires the colliding winds to be dense and relatively slow.
Note also that gamma-ray emission in hadronic models
is not significantly affected by
synchrotron losses, as the gamma-rays are produced predominantly via $\pi_0$ decay.
On the other hand, the general lack of observed hard X-rays (\citealt{deBecker+07})
could be explained in the hadronic scenario,
owing to the abovementioned suppression via synchrotron and bremsstrahlung losses.

\acknowledgements

We thank Andrei M. Beloborodov, Guillaume Dubus, Pierrick Martin, and Koji Mukai for helpful conversations.  BDM gratefully acknowledges support from NASA grants NNX15AU77G (Fermi),
NNX15AR47G, NNX16AB30G (Swift), and NNX16AB30G (ATP), NSF grant AST-1410950, the Alfred P. Sloan Foundation, and the Research Corporation for Science Advancement through the Scialog Program (Grant number RCSA 23810).

\bibliographystyle{aasjournal}

\end{document}